\documentclass[english,10pt,final,journal,letterpaper,onecolumn,twoside]{IEEEtran}
\usepackage[T1]{fontenc}
\usepackage[latin9]{inputenc}
\usepackage{geometry}
\usepackage{float}
\usepackage{mathrsfs}
\usepackage{amsthm}
\usepackage{amsmath}
\usepackage{amssymb}
\usepackage{graphicx}
\usepackage{esint}
\usepackage{hyperref}

\makeatletter

\floatstyle{ruled}
\newfloat{algorithm}{tbp}{loa}
\providecommand{\algorithmname}{Algorithm}
\floatname{algorithm}{\protect\algorithmname}

  \theoremstyle{remark}
  \newtheorem{rem}{\protect\remarkname}
  \theoremstyle{definition}
  \newtheorem{problem}{\protect\problemname}
  \theoremstyle{definition}
  \newtheorem{defn}{\protect\definitionname}
  \theoremstyle{plain}
  \newtheorem{thm}{\protect\theoremname}

\IEEEoverridecommandlockouts

\usepackage{dsfont}
\usepackage{relsize}
\usepackage{subdepth}
\allowdisplaybreaks

\author{
    \IEEEauthorblockN{Dionysios S. Kalogerias, \IEEEmembership{Student Member, IEEE}
            \\and Athina P. Petropulu$^*$, \IEEEmembership{Fellow, IEEE}} 
\thanks{D. S. Kalogerias and A. P. Petropulu are with the Department of Electrical \& Computer Engineering, Rutgers, The State University of New Jersey, 94 Brett Rd, Piscataway, 08854, NJ, USA. e-mail: \{d.kalogerias, athinap\}@rutgers.edu}
    \thanks{This work is supported by the National Science Foundation (NSF) under grant CNS - 1239188.}
}

\usepackage{dsfont}

\usepackage{relsize}
\usepackage{subdepth}

\providecommand{\definitionname}{Definition}

\providecommand{\theoremname}{Theorem}

\providecommand{\definitionname}{Definition}

\providecommand{\remarkname}{Remark}
\providecommand{\theoremname}{Theorem}

\usepackage[english]{babel}
\usepackage{url}

\DeclareFontFamily{OT1}{pzc}{}
\DeclareFontShape{OT1}{pzc}{m}{it}{<-> s * [1.200] pzcmi7t}{}
\DeclareMathAlphabet{\mathpzc}{OT1}{pzc}{m}{it}

\usepackage[noadjust]{cite}
\usepackage{filecontents}

\usepackage{mathtools}

\usepackage{kerkis}

\usepackage{changepage}

\def\section{\@startsection{section}{1}{\z@}{3.0ex plus 1.5ex minus 1.5ex}
{0.7ex plus 1ex minus 0ex}{\normalfont\large\centering\scshape}}%

\@ifundefined{showcaptionsetup}{}{%
 \PassOptionsToPackage{caption=false}{subfig}}
\usepackage{subfig}

\makeatother

\providecommand{\definitionname}{Definition}
\providecommand{\problemname}{Problem}
\providecommand{\remarkname}{Remark}
\providecommand{\theoremname}{Theorem}

\begin{document}

\title{\begin{adjustwidth}{-1cm}{-1cm} \begin{center}Sequential Channel
State Tracking \& SpatioTemporal Channel Prediction in\\
Mobile Wireless Sensor Networks\end{center} \end{adjustwidth}}
\maketitle
\begin{abstract}
\begin{adjustwidth}{1cm}{1cm}We propose a nonlinear filtering framework
for approaching the problems of channel state tracking and spatiotemporal
channel gain prediction in mobile wireless sensor networks, in a Bayesian
setting. We assume that the wireless channel constitutes an observable
(by the sensors/network nodes), spatiotemporal, conditionally Gaussian
stochastic process, which is statistically dependent on a set of \textit{hidden}
channel parameters, called the \textit{channel state}. The channel
state evolves in time according to a known, \textit{non stationary,
nonlinear and/or non Gaussian} Markov stochastic kernel. This formulation
results in a partially observable system, with a temporally varying
global state and spatiotemporally varying observations. Recognizing
the intractability of general nonlinear state estimation, we advocate
the use of grid based approximate filters as an effective and robust
means for recursive tracking of the channel state. We also propose
a sequential spatiotemporal predictor for tracking the channel gains
at any point in time and space, providing real time sequential estimates
for the respective channel gain map, for each sensor in the network.
Additionally, we show that \textit{both} estimators converge towards
the true respective MMSE optimal estimators, in a common, relatively
strong sense. Numerical simulations corroborate the practical effectiveness
of the proposed approach.\end{adjustwidth}\end{abstract}


\setlength{\textfloatsep}{10pt}

\section{Introduction}

\IEEEPARstart{A}s a result of the growing interest in wireless
networks and distributed communication and processing systems, new,
challenging problems have recently transpired, related not only to
the flow of information over networks, but also to the estimation
and control of the underlying physical layer. In a large number of
important applications, accurate estimation of Channel State Information
(CSI), or Statistical CSI (SCSI), for all of the nodes/sensors in
a wireless network is essential. Popular examples include distributed
collaborative beamforming and related Space Division Multiple Access
(SDMA) techniques, target detection and estimation in distributed
networked radar systems, and information theoretic physical layer
security via transmission optimization, just to name a few \cite{LiPetropuluPoor2011,NikosBeam-1,Petropulu_1_2010,Petropulu_2_2008,Petropulu_3_2009,Petropulu_4_2009}.

Traditionally, in such applications, CSI and SCSI estimation is done
via pilot based schemes, blind channel estimation techniques, or even
through averaging from rough channel observations. Except for the
fact that naive extension of the conventional techniques to larger
scale wireless networks requires collaboration between the network
nodes, which can be both bandwidth and power intensive, these techniques
are only sufficient for relatively lower rate and/or quasistatic environments,
where the statistics of the communication medium do not change significantly
over time. However, the behavior of most indoor and outdoor communication
environments of practical interest is intrinsically time varying (see,
e.g., \cite{Trappe2_2009}).

In addition to the temporal variation of the wireless medium, recently,
considerable interest has been expressed concerning its spatial variation
as well. In fact, learning how the communication channel evolves through
space is tightly connected to the ability of a network to assess the
quality of the channel at previously unexplored locations in the space,
based on local channel measurements at the respective sensors and
by exploiting spatial statistical correlations among them. Such knowledge
would be beneficial in a number of new and important applications.
Examples include mobile beamforming \cite{CLPZHawaii2012}, mobility
enhanced physical layer security, \cite{Trappe3_2008,KalPet-Jammers-2013,KalPet-Mobi-2014},
communication-aware motion and path planning, network routing, connectivity
maintenance and physical layer based dynamic coverage \cite{Mostofi_3_2011,Mostofi_4_2012,Mostofi_5_2014}.
In all these cases, \textit{dynamic spatiotemporal channel estimation/tracking
and prediction} becomes an essential part of mobility control, since
it would provide valuable physical layer related information (channel
maps), which is absolutely necessary for dynamic decision making and
stochastic control.

Regarding the explicit use of the idea of \textit{parameter tracking}
in channel estimation, important work has been done on identification/characterization
of multipath wireless channels. For example, in \cite{Multipath_1_2011},
a sparse variational Bayesian extension of the popular SAGE algorithm
\cite{EM_SAGE_1994} was developed, aiming to high resolution parameter
estimation of the multipath components of spatially and frequency
selective wireless channels. In \cite{Multipath_2_2009}, the problems
of detection, estimation and tracking of MIMO radio propagation parameters
were considered, where an efficient state space approach was developed,
based on the proper use of the Extended Kalman Filter. A similar problem
was also considered in \cite{Multipath_3_2008}, where a specially
designed estimation algorithm was proposed, based on particle filtering.

To the best of the authors' knowledge, the first basic approach to
joint spatiotemporal channel (specifically shadowing) tracking and
prediction was recently presented in \cite{Giannakis_Spatial1_2011,Giannakis_Spatial2_2011},
where the use of Channel Gain (CG) maps was advocated as an advantageous
alternative to Power Spectral Density (PSD) maps for cooperative spectrum
sensing in the context of cognitive radios. The overall formulation
of the problem presented in \cite{Giannakis_Spatial1_2011,Giannakis_Spatial2_2011}
is based on a direct fusion of previously proposed results in wireless
channel modeling \cite{Channel_Modeling1_2009} and spatiotemporal
Kalman filtering \cite{KKF_1_1999}, also known in the literature
as Kriged Kalman Filtering (KKF) \cite{KKF_2_1998,KKF_3_2009}. Although
analytically appealing, the state space model considered in \cite{Giannakis_Spatial1_2011,Giannakis_Spatial2_2011}
for describing the spatiotemporal evolution of the wireless channel
is rather restrictive; both the dependence of the shadowing field
on its previous value in time and its spatial interactions are characterized
by purely linear functional relationships\textit{, }focusing mainly
on\textit{ }modeling the spatiotemporal variations of the\textit{
trend }of the field.

In this work, in order to facilitate conceptualization, we consider
a simple network configuration, comprised by a ``reference'' point/antenna
capable of broadcasting global information in the space, as well as
a set of \textit{possibly mobile} network nodes/sensors, capable either
of local message exchange, or communicating with a fusion center (see
Fig. \ref{fig:Network}). For concreteness, a channel learning scenario
is considered where, at each time instant, the reference antenna broadcasts
a generic signal to all the sensors at the same time, which in turn
use the acquired measurements in order to learn the basic characteristics
of the channel, and subsequently make consistent predictions regarding
its quality for any point in time and space. The statistical model
describing the joint spatiotemporal behavior of the channel measurements
gathered at the sensors is inspired by \cite{MostofiSpatial2012}.
However, different from \cite{MostofiSpatial2012}, the descriptive
channel parameters (e.g., the path loss exponent, the shadowing power,
etc.), referred to here as the \textit{channel state}, are assumed
to be temporally varying. Specifically, we assume that the whole channel
state constitutes a Markov process, with known, but potentially \textit{non
stationary, nonlinear and/or non Gaussian} transition model. Essentially,
under this formulation, the spatiotemporal evolution of the channel
is conveniently modeled as a general two layer stochastic system,
or, in more specific terms, as a partially observable dynamical system
(with Markovian dynamics), more commonly referred to as a Hidden Markov
Model (HMM) \cite{Elliott1994Hidden}.

The proposed formulation can naturally lead to a \textit{full blown}
state space channel description in terms of generality. Compared to
\cite{Giannakis_Spatial1_2011}, it is more general, since it can
deal with complex variations in the channel characteristics, other
than linear variations in the shadowing trend. However, in our state
space description of the channel, spatial statistical dependencies
are present only in the observations process, whereas in \cite{Giannakis_Spatial1_2011},
the trend of the shadowing component of the channel, constituting
the hidden state, respectively, is \textit{jointly} spatiotemporally
colored. Also, here, we will consider the \textit{detrended} problem,
similar to the one treated in \cite{MostofiSpatial2012} (in a non
Bayesian framework) and which has proven to be in good agreement with
reality as well. A complete channel model, combining both a non zero
spatiotemporally varying shadowing trend in the fashion of \cite{Giannakis_Spatial1_2011,Giannakis_Spatial2_2011}
with temporally varying channel parameters advocated here, results
in a non trivial problem in nonlinear estimation and constitutes a
subject of future research.

Our main contributions are clear and summarized in the following.
\textbf{1)} Recognizing the obvious intractability of state estimation
in partially observable nonlinear systems, we propose the use of grid
based approximate nonlinear recursive filters for \textit{sequential
channel state tracking}. Due to the relatively small dimension of
the channel state, grid based methods constitute excellent approximation
candidates for the problems at hand. Then, exploiting filtered estimates
of the channel state, a recursive spatiotemporal predictor of the
channel gains (magnitudes) is developed, providing \textit{real time
sequential estimates} for the respective CG map, for each sensor in
the network.\textbf{ 2)} We provide a set of simple, relaxed conditions,
under which the proposed channel state tracker briefly described above
is \textit{asymptotically optimal}, in the sense that it converges
to the respective true MMSE channel state estimator, in a relatively
strong sense. The convergence of the proposed spatiotemporal predictor
is established in exactly the same sense, providing a unified convergence
criterion for both sequential estimators.

The results presented in this paper essentially show that grid based
approximate nonlinear filtering is meaningfully applicable to the
channel state tracking and spatiotemporal channel prediction problems
of interest. As we will see, this is possible by approximating the
complex nonlinearly varying processes modeling the evolution of the
channel parameters by appropriately designed Markov chains with finite
state spaces. And in the other way around, the asymptotic optimality
properties of the proposed approach clearly justify the use of such
Markov chains as an excellent approximation choice for the highly
nonlinear problems at hand. 

The paper is organized as follows. In Section II, we present a detailed
formulation of the problems of interest, along with some mild technical
assumptions on the structure of the HMM channel description under
consideration. In Section III, we present a number of essential results
on asymptotically optimal, grid based recursive filtering and prediction
of hidden Markov processes. Section IV is devoted to the development
of the proposed channel state tracking and spatiotemporal channel
prediction schemes, along with a complete theoretical justification,
including the presentation of new asymptotic results. In Section V,
representative numerical simulations are presented, corroborating
the practical effectiveness of the proposed approach. Finally, Section
VI concludes the paper.

\section{System Model \& Problem Formulation}

\begin{figure}
\centering\includegraphics[scale=0.9]{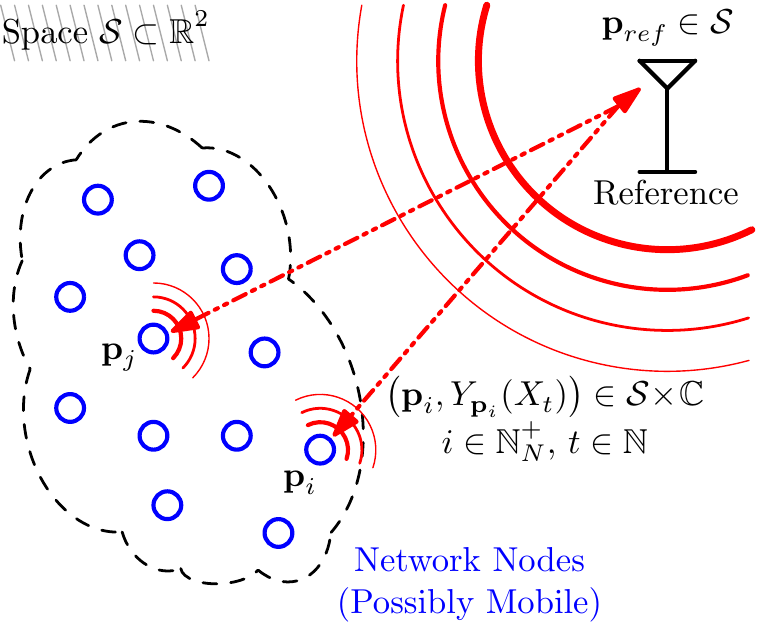}

\protect\caption{\label{fig:Network}The wireless network of interest.}
\end{figure}

For simplicity, we consider a wireless network of typical form, a
high level illustration of which is shown in Fig. \ref{fig:Network}.
The environment is assumed to be a closed planar region ${\cal S}\subset\mathbb{R}^{2}$,
where, as already stated above, there exists a fixed, stationary antenna
at a reference position, capable of at least information broadcasting.
There also exist a set of $N$ single antenna sensors, possibly mobile,
monitoring the channel relative to the reference antenna. These sensors
may be a subset of the total nodes in the network and are responsible
for the respective channel estimation tasks. The sensors can cooperate,
and further, can either communicate with a fusion center (in a centralized
setting), or exchange basic messages amongst each other (in a decentralized/infrastructureless
scenario) using a low rate dedicated channel. Concerning channel modeling,
we adopt a flat fading model between each node and the reference antenna.
It is additionally assumed that channel reciprocity holds and that
all network nodes can perfectly observe their individual channel realizations
(e.g. magnitudes and potentially phases) relative to the reference
antenna \cite{MostofiSpatial2012}. The channels are modeled as spatially
and temporally correlated, discrete time random processes (spatiotemporal
random fields), sharing the same channel environment, at least as
far as the underlying characteristics of the communication medium
are concerned.

As already mentioned, the channel state encompasses statistics of
the communication medium, and is here modeled as a multidimensional
discrete time stochastic process, evolving in time according to a
known statistical model. The channel state is assumed to be hidden
from the network nodes; the nodes can observe their respective channel
realizations, but they cannot directly observe the characteristics
of the mechanism that generates these realizations. Naturally, this
stochastic structure gives rise to the description of the channel
state and the associated observation process(es) as a \textit{partially
observable stochastic dynamical system.}

Let us describe our problem in more explicit mathematical terms. All
stochastic processes are defined on a common probability triplet $\left(\Omega,\mathscr{F},{\cal P}\right)$.
Let $X_{t}\equiv X_{t}\left(\omega\right)\subset\mathbb{R}^{M\times1}$,
$t\in\mathbb{N}$, $\omega\in\Omega$ denote the hidden channel state.
Under the flat fading assumption, the relative to the reference antenna
complex channel process at each network node $i\in\mathbb{N}_{N}^{+}$,
located at ${\bf p}_{i}\equiv{\bf p}_{i}\left(t\right)\in{\cal S}$
(that is, the nodes might be moving), can be decomposed as \cite{Goldsmith2005Wireless}
\begin{align}
\mathbb{C} & \ni Y_{i}\left({\bf p}_{i}\left(t\right),X_{t}\right)\equiv Y_{{\bf p}_{i}}\left(X_{t}\right)\equiv\underbrace{Y_{{\bf p}_{i}}^{PL}\left(X_{t}\right)}_{\text{path loss}}\underbrace{Y_{i}^{SH}\left(X_{t}\right)}_{\text{shadowing}}\underbrace{Y_{{\bf p}_{i}}^{MF}\left(t\right)\exp\left(\mathfrak{J}\dfrac{2\pi d_{i}\left(t\right)}{\lambda}\right)}_{\text{fading}},\label{eq:Channel_1}
\end{align}
where $\mathfrak{J}\triangleq\sqrt{-1}$, $\lambda\in\mathbb{R}_{++}$
denotes the wavelength employed for the communication and where: \textbf{1)
}$Y_{{\bf p}_{i}}^{PL}\left(X_{t}\right)\in\mathbb{R}$ denotes path
loss, defined as $Y_{{\bf p}_{i}}^{PL}\left(X_{t}\right)\triangleq\left\Vert {\bf p}_{i}\left(t\right)-{\bf p}_{ref}\right\Vert _{2}^{-\mu\left(X_{t}\right)/2}\triangleq\left(d_{i}\left(t\right)\right)^{-\mu\left(X_{t}\right)/2}$,
where $\mu\left(X_{t}\right)\in\mathbb{R}_{++}$ is the state dependent
path loss exponent, which is the same for all network nodes and ${\bf p}_{ref}\in{\cal S}$
denotes the position of the reference antenna in ${\cal S}$. \textbf{2)}
$Y_{i}^{SH}\left(X_{t}\right)\in\mathbb{R}$ denotes the shadowing
part of the channel model and its square, \textit{conditionally on
}$X_{t}$, constitutes a base-$10$ log-normal random variable with
zero location and scale \textit{depending on} $X_{t}$. \textbf{3)
}$Y_{{\bf p}_{i}}^{MF}\left(t\right)\in\mathbb{C}$ represents multipath
fading, which, for simplicity, is assumed to be a \textit{spatiotemporally}
white%
\footnote{See \cite{MostofiSpatial2012} and references therein for arguing
about the validity of this assumption. Also, throughout the paper,
the samples of a discrete time white stochastic process are understood
to be independent. %
}, strictly stationary process with fully known statistical description,
not associated with $X_{t}$, therefore being an unpredictable complex
``observation noise''. Making the substitution $Y_{i}\left({\bf p}_{i}\left(t\right),X_{t}\right)\leftarrow$$\exp\left(-\mathfrak{J}2\pi d_{i}\left(t\right)/\lambda\right)$$Y_{i}\left({\bf p}_{i}\left(t\right),X_{t}\right)$
and using properties of the complex logarithm, we can define the observations
of the $i$-th node in logarithmic scale as
\begin{multline}
y_{t}^{i}\triangleq10\log_{10}\left|Y_{{\bf p}_{i}}\left(X_{t}\right)\right|^{2}-10\mathbb{E}\left\{ \log_{10}\left|Y_{{\bf p}_{i}}^{MF}\left(t\right)\right|^{2}\right\} =\\
-10\mu\left(X_{t}\right)\log_{10}\left(d_{i}\left(t\right)\right)+10\log_{10}\left(Y_{i}^{SH}\left(X_{t}\right)\right)^{2}+\\
+\overline{10\log_{10}\left|Y_{{\bf p}_{i}}^{MF}\left(t\right)\right|^{2}}\triangleq\alpha_{t}^{i}\mu\left(X_{t}\right)+\sigma_{t}^{i}\left(X_{t}\right)+\xi_{t}^{i},\label{eq:Amplitude_log}
\end{multline}
where $\overline{\left(\cdot\right)}$ denotes the zero mean version
of a random variable. We should emphasize here that by ``measurement''
or ``observation'' we refer to the \textit{predictable} component
of the channel, which is described in terms of the channel magnitude.

In similar fashion as in \cite{MostofiSpatial2012}, the following
further assumptions are made%
\footnote{In what follows, ``i.d.'' means ``identically distributed'' and
``i.i.d.'' means ``independent and i.d.''.%
}: $\xi_{t}^{i}\overset{i.i.d.}{\sim}{\cal N}\left(0,\sigma_{\xi}^{2}\right),$
$\forall i\in\mathbb{N}_{N}^{+}$ and $\forall t\in\mathbb{N}$. This
is a simplified, although quite reasonable assumption. Also see \cite{Cotton2007}.
Second, conditional on $X_{t}$, $\sigma_{t}^{i}\left(X_{t}\right)\overset{i.d.}{\sim}{\cal N}\left(0,\eta^{2}\left(X_{t}\right)\right),\forall i\in\mathbb{N}_{N}^{+}$.
This stems from the fact that $\left(Y_{i}^{SH}\left(X_{t}\right)\right)^{2}$
is (base-$10$) log-normally distributed. Additionally, for each set
of positions of the network nodes in ${\cal S}$, it is assumed that
the members of the set $\left\{ \sigma_{t}^{i}\left(X_{t}\right)\right\} _{i\in\mathbb{N}_{N}^{+}}$
constitute \textit{jointly normal, spatially correlated} random variables
with (symmetric and positive definite in $\left({\bf p}_{i}\left(t\right),{\bf p}_{j}\left(t\right)\right)$)
conditional on $X_{t}$ autocorrelation kernel
\begin{equation}
{\cal R}\left({\bf p}_{i}\left(t\right),{\bf p}_{j}\left(t\right),\boldsymbol{\theta}\left(X_{t}\right)\right):\left({\cal S}\times{\cal S}\times\mathbb{R}^{K}\right)\mapsto\mathbb{R},
\end{equation}
for $\left(i,j\right)\in\mathbb{N}_{N}^{+}\times\mathbb{N}_{N}^{+}$,
where $K$ denotes the dimension of the state dependent parameter
vector $\boldsymbol{\theta}\left(X_{t}\right)$, with
\begin{flalign}
{\cal R}\left({\bf p}_{i}\left(t\right),{\bf p}_{i}\left(t\right),\boldsymbol{\theta}\left(X_{t}\right)\right) & \equiv\eta^{2}\left(X_{t}\right),\quad\forall i\in\mathbb{N}_{N}^{+}.
\end{flalign}
Therefore, the $\left(i.j\right)$-th entry of the time evolving,
\textit{conditional (on $X_{t}$) covariance} matrix of the random
vector $\boldsymbol{\sigma}_{t}\left(X_{t}\right)\triangleq\left[\left\{ \sigma_{t}^{i}\left(X_{t}\right)\right\} _{i\in\mathbb{N}_{N}^{+}}\right]^{\boldsymbol{T}}\in\mathbb{R}^{N\times1}$,
$\boldsymbol{\Sigma}_{t}\left(\boldsymbol{\theta}\left(X_{t}\right)\right)\equiv\mathbb{E}\left\{ \left.\boldsymbol{\sigma}_{t}\left(X_{t}\right)\left(\boldsymbol{\sigma}_{t}\left(X_{t}\right)\right)^{\boldsymbol{T}}\right|X_{t}\right\} $
$\in{\cal D}_{\boldsymbol{\Sigma}}\subset\mathbb{R}^{N\times N}$,
with ${\cal D}_{\boldsymbol{\Sigma}}$ bounded, is defined as
\begin{flalign}
\boldsymbol{\Sigma}_{t}\left(\boldsymbol{\theta}\left(X_{t}\right)\right) & \left(i,j\right)\triangleq{\cal R}\left({\bf p}_{i}\left(t\right),{\bf p}_{j}\left(t\right),\boldsymbol{\theta}\left(X_{t}\right)\right),
\end{flalign}
$\forall\left(i,j\right)\in\mathbb{N}_{N}^{+}\times\mathbb{N}_{N}^{+}$.
For instance, for the heat kernel type of autocorrelation function
employed in \cite{MostofiSpatial2012} and proposed much earlier in
\cite{Gudmundson1991}, defined as
\begin{align}
{\cal R}\left({\bf p}_{i}\left(t\right),{\bf p}_{j}\left(t\right),\boldsymbol{\theta}\left(X_{t}\right)\right) & \triangleq{\cal R}\left(d_{ij}\left(t\right),\boldsymbol{\theta}\left(X_{t}\right)\right)\triangleq\theta_{1}\left(X_{t}\right)\exp\left(-\dfrac{d_{ij}\left(t\right)}{\theta_{2}\left(X_{t}\right)}\right),\label{eq:isotropic}
\end{align}
where $d_{ij}\left(t\right)\triangleq\left\Vert {\bf p}_{i}\left(t\right)-{\bf p}_{j}\left(t\right)\right\Vert _{2}\in\mathbb{R}_{+}$,
we make the identifications $\boldsymbol{\theta}\left(X_{t}\right)\triangleq\left[\theta_{1}\left(X_{t}\right)\,\theta_{2}\left(X_{t}\right)\right]^{\boldsymbol{T}}$
with $\theta_{1}\left(X_{t}\right)\equiv\eta^{2}\left(X_{t}\right)$.
The first parameter, $\theta_{1}\left(X_{t}\right)$, called the \textit{shadowing
power}, controls the variance of the shadowing part of the channel,
whereas the second, $\theta_{1}\left(X_{t}\right)$, called the \textit{correlation
distance}, controls the decay rate of the spatial correlation between
the channels for each pair of network nodes. This simple isotropic
model will be employed in the numerical simulations presented in Section
V.

In order to completely define an overall observation process for all
nodes in the network, we may stack the $N$ individual channel processes
of \eqref{eq:Amplitude_log}, resulting in the vector additive observation
model
\begin{equation}
{\bf y}_{t}\equiv\boldsymbol{\alpha}_{t}\mu\left(X_{t}\right)+\boldsymbol{\sigma}_{t}\left(X_{t}\right)+\boldsymbol{\xi}_{t},\quad\forall t\in\mathbb{N},\label{eq:Observations_vector}
\end{equation}
where $\boldsymbol{\sigma}_{t}\left(X_{t}\right)$ is defined as above
and ${\bf y}_{t}\in\mathbb{R}^{N\times1}$, $\boldsymbol{\alpha}_{t}\in\mathbb{R}^{N\times1}$
and $\boldsymbol{\xi}_{t}\in\mathbb{R}^{N\times1}$ are defined accordingly.
The observation process \eqref{eq:Observations_vector} can also be
rewritten in the canonical form ${\bf y}_{t}\equiv\boldsymbol{\alpha}_{t}\mu\left(X_{t}\right)+\sqrt{{\bf C}_{t}\left(X_{t}\right)}\boldsymbol{u}_{t},\forall t\in\mathbb{N},$
where $\boldsymbol{u}_{t}\equiv\boldsymbol{u}_{t}\left(\omega\right)$
constitutes a standard Gaussian white noise process and ${\bf C}_{t}\left(X_{t}\right)\triangleq\boldsymbol{\Sigma}_{t}\left(\boldsymbol{\theta}\left(X_{t}\right)\right)+\sigma_{\xi}^{2}{\bf I}_{N\times N}\in{\cal D}_{{\bf C}}$,
with ${\cal D}_{{\bf C}}$ obviously bounded.

Let us now concentrate more on the time evolving underlying channel
state process $X_{t}\in\mathbb{R}^{M\times1}$. In this work, we will
assume that $X_{t}$ constitutes a Markov process with \textit{known
but nonlinear and (possibly) nonstationary dynamics}, described by
a possibly nonstationary stochastic kernel%
\footnote{Throughout the paper, we use the intuitive notation ${\cal K}_{t}\left({\cal A}\left|X_{t-1}\left(\omega\right)\equiv\boldsymbol{x}\right.\right)\equiv{\cal K}_{t}\left(\left.{\cal A}\right|\boldsymbol{x}\right)$,
for ${\cal A}$ Borel.%
} ${\cal K}_{t}:\mathscr{B}\left(\mathbb{R}^{M\times1}\right)\times\mathbb{R}^{M\times1}\mapsto\left[0,1\right]$.
Also, we will make the generic and realistic assumption that the state
is confined to a compact strict subset of $\mathbb{R}^{M\times1}$,
that is, $\forall t\in\mathbb{N},X_{t}\in\left[a,b\right]^{M}\triangleq{\cal Z}\subset\mathbb{R}^{M\times1}$,
almost surely. Depending on the available information, instead of
using stochastic kernels, we may alternatively assume the existence
of an explicit state transition model expressing the temporal evolution
of the state, defined as $X_{t}\triangleq f_{t}\left(X_{t-1},W_{t}\right)\in{\cal Z},\forall t\in\mathbb{N},$
where, for each $t$, $f_{t}:{\cal Z}\times{\cal W}\overset{a.s.}{\mapsto}{\cal Z}$
constitutes a measurable nonlinear state transition mapping with somewhat
``favorable'' analytical behavior (see below) and $W_{t}\equiv W_{t}\left(\omega\right)\in{\cal W}\subseteq\mathbb{R}^{M_{W}\times1}$,
for $t\in\mathbb{N}$, $\omega\in\Omega$, denotes a (discrete time)
white noise process with known measure and state space ${\cal W}$.

From now on, in order to facilitate the presentation and without loss
of generality, we will drop the subscript ``$t$'' both in the stochastic
kernels and transition mappings governing $X_{t}$, therefore assuming
stationarity of the state. Further, for mathematical simplicity, although
there are endless possibilities for defining the state dependent functions
$\mu\left(X_{t}\right)$ and $\boldsymbol{\theta}\left(X_{t}\right)$,
we will assume that $\mu\left(X_{t}\right)\equiv X_{t}\left(1\right)\in\mathbb{R}$
and $\boldsymbol{\theta}\left(X_{t}\right)\equiv\left[X_{t}\left(2\right)\,\ldots\, X_{t}\left(M\right)\right]^{\boldsymbol{T}}\in\mathbb{R}^{\left(M-1\right)\times1}$,
also in agreement with our intuition. From the previous discussion,
it follows that the partially observable system defined above constitutes
a HMM and can be equivalently described by the system of stochastic
difference equations
\begin{equation}
\begin{array}{l}
\begin{cases}
X_{t}\left|X_{t-1}\right.\sim{\cal K}\left(X_{t}\in\text{d}x\left|X_{t-1}\right.\right) & \text{or}\\
X_{t}\equiv f\left(X_{t-1},W_{t}\right)
\end{cases}\\
\,\hspace{0.86em}{\bf y}_{t}\equiv\boldsymbol{A}_{t}X_{t}+\boldsymbol{\sigma}_{t}\left(X_{t}\right)+\boldsymbol{\xi}_{t}
\end{array},\enskip\forall t\in\mathbb{N},\label{eq:State_Space-1}
\end{equation}
where $\boldsymbol{A}_{t}\triangleq\left[\boldsymbol{\alpha}_{t}\:{\bf 0}_{N\times\left(M-1\right)}\right]\in\mathbb{R}^{N\times M}$.
In addition to the above and in favor of supporting our analytical
arguments presented in subsequent sections, we make the following
mild assumptions on the functional structure of the observation process
of the HMM described by \eqref{eq:State_Space-1}.\textbf{\medskip{}
}

\noindent \textbf{Assumption 1:} \textbf{(Continuity \& Expansiveness)}
All members of the functional family\linebreak{}
$\left\{ \boldsymbol{\Sigma}_{t}:{\cal Z}\mapsto{\cal D}_{\boldsymbol{\Sigma}}\right\} _{t\in\mathbb{N}}$
are \textit{elementwise} uniformly Lipschitz continuous, that is,
there exists some universal and bounded constant $K_{\boldsymbol{\Sigma}}\in\mathbb{R}_{+}$,
such that, $\forall t\in\mathbb{N}$ and $\forall\left(i,j\right)\in\mathbb{N}_{N}^{+}\times\mathbb{N}_{N}^{+}$,
\begin{equation}
\left|\boldsymbol{\Sigma}_{t}^{ij}\left(\boldsymbol{x}\right)-\boldsymbol{\Sigma}_{t}^{ij}\left(\boldsymbol{y}\right)\right|\le K_{\boldsymbol{\Sigma}}\left\Vert \boldsymbol{x}-\boldsymbol{y}\right\Vert _{1},
\end{equation}
$\forall\left(\boldsymbol{x},\boldsymbol{y}\right)\in{\cal Z}\times{\cal Z}$.
If $\boldsymbol{x}$ is substituted by the stochastic process $X_{t}\left(\omega\right)$,
then all the above statements continue to hold almost surely. Also,
it is true that
\begin{equation}
\lambda_{inf}\equiv{\displaystyle \inf_{t\in\mathbb{N}}}\hspace{0.2em}{\displaystyle \inf_{\boldsymbol{x}\in{\cal Z}}}\hspace{0.2em}\lambda_{min}\left({\bf C}_{t}\left(\boldsymbol{x}\right)\right)>1,
\end{equation}
a requirement which can always be satisfied by appropriate normalization
of the observations.\textbf{\medskip{}
}

For later reference (Section V), let us note that the isotropic autocorrelation
kernel previously defined by \eqref{eq:isotropic} can be very easily
verified to satisfy the Lipschitz condition of Assumption 1, simply
considering the compactness of the state vector.
\begin{rem}
The assumption of $X_{t}$ satisfying the (first order) Markov property
does not offer only analytical tractability, but also practical feasibility.
For example, statistical inference in higher order HMMs suffer from
the curse of dimensionality so much that, most of the times, the computational
effort required for the implementation of basic state estimators is
absolutely prohibitive. On the other hand, our proposed formulation
is based on general nonlinear models for describing the statistical
behavior of the state, offering far greater flexibility, as well as
modeling precision, compared to classical linear difference equations.
From another point of view, it is well known that a tremendous amount
of real world dynamical systems can be modeled using Markov processes
and that, in cases where this is not entirely true, Markov processes
usually constitute very good modeling approximations.
\end{rem}
Let us now define the problems of interest in this paper in a mathematically
precise way. Hereafter, \textit{strict optimality} will be meant to
be in the \textit{Minimum Mean Square Sense (MMSE)}. Also, in the
following, the natural filtration generated by the causal observation
process ${\bf y}_{t}$ is defined as 
\begin{equation}
\left\{ \mathscr{Y}_{t}\right\} _{t\in\mathbb{N}}\triangleq\left\{ \sigma\left\{ \left\{ {\bf y}_{i}\right\} _{i\in\mathbb{N}_{t}}\right\} \right\} _{t\in\mathbb{N}},
\end{equation}
where $\sigma\left\{ A\right\} $ denotes the $\sigma$-algebra generated
by the random element $A$.
\begin{problem}
\textbf{(Sequential Channel State Tracking (SCST))} \textit{Develop
a theoretically grounded, sequential scheme for (approximately) evaluating
the strictly optimal filter or $\rho$-step predictor of the channel
state $X_{t}$ on the basis of the available channel magnitude observations
up to time $t$, given by
\begin{equation}
\widehat{X}_{t+\rho}\triangleq\mathbb{E}\left\{ \left.X_{t+\rho}\right|\mathscr{Y}_{t}\right\} ,\quad\forall t\in\mathbb{N},
\end{equation}
where $\rho\ge0$ constitutes the }\textbf{\textit{prediction horizon}}\textit{.
The computational complexity of the sequential scheme may not grow
as more observations become available.}
\end{problem}

\begin{problem}
\textbf{(Sequential Spatiotemporal Channel Prediction (SSCP))} \textit{Develop
a theoretically grounded, sequential scheme for (approximately) evaluating
the strictly optimal spatiotemporal predictor of the channel magnitude
at position ${\bf q}\in\mathbb{R}^{2}$ and time $t+\rho$ ($\rho\ge0$
is the prediction horizon) given the available channel magnitude observations
up to time $t$, expressed as
\begin{equation}
\widehat{y}_{t+\rho}\left({\bf q}\right)\triangleq\mathbb{E}\left\{ \left.y_{t+\rho}\left({\bf q}\right)\right|\mathscr{Y}_{t}\right\} ,\quad\forall t\in\mathbb{N}.
\end{equation}
Again, the computational complexity of the sequential scheme may not
grow as more observations become available.}\end{problem}
\begin{rem}
The SSCP problem is clearly related to the channel predictability
framework of \cite{MostofiSpatial2012}. In fact, in this paper, we
use almost the same channel description (observation process). However,
our proposed framework is philosophically different and potentially
more general than that proposed in \cite{MostofiSpatial2012}, since
the underlying channel dynamics (the channel state) are time varying
and the considered estimation and prediction problems are formulated
in a \textit{Bayesian sense}.
\end{rem}
As we will see in later in Section IV, the SSCP problem can be solved
sequentially using the respective sequential solution of the SCST
problem. However, unfortunately, it is well known that, except for
some very special cases such as those where the state process $X_{t}$
satisfies a linear recursion or where it constitutes a Markov chain
(discrete state space) \cite{Segall_Point1976,Marcus1979,Elliott1994Exact,Elliott1994_HowToCount},
the respective nonlinear filtering and prediction problems do not
admit any known sequential (in particular, recursive) representation
\cite{Segall1976,Elliott1994Hidden}. Therefore, in order to solve
the SCST problem defined above, one typically has to rely on carefully
designed and robust approximations to the problem of nonlinear filtering
of Markov processes in discrete time, focusing on the class of systems
(HMMs) described by \eqref{eq:State_Space-1}. This is exactly the
subject of the next section.

\section{\label{sec:Quant_Filtering}Asymptotically Optimal Recursive Filtering
\& Prediction of Markov Processes: Prior Results \& Preliminaries}

In the following, we present a number of important results in asymptotically
optimal, approximate recursive filtering of Markov processes, recently
presented in \cite{KalPetGRID2014}. These results will provide us
with the required mathematical tools for attacking the SCST and SSCP
problems, defined previously in Section II.

\subsection{Uniform State Quantizations}

From Section II, we have assumed that $X_{t}\in{\cal Z}\equiv\left[a,b\right]^{M},\forall t\in\mathbb{N},a.s.,$
where, geometrically, ${\cal Z}$ constitutes an $M$-hypercube, representing
the compact set of support of the state $X_{t}$. Let us discretize
${\cal Z}$ into $L_{S}\triangleq L^{M}$ hypercubic $M$-dimensional
cells of identical volume (each dimension is partitioned in to $L$
intervals). The center of mass of the $l$-th cell is denoted as $\boldsymbol{x}_{L_{S}}^{l},l\in\mathbb{N}_{L_{S}}^{+}$.
Then, letting ${\cal X}_{L_{S}}\triangleq\left\{ \boldsymbol{x}_{L_{S}}^{l}\right\} _{l\in\mathbb{N}_{L_{S}}^{+}}$,
the \textit{quantizer} ${\cal Q}_{L_{S}}:\left({\cal Z},\mathscr{B}\left({\cal Z}\right)\right)\mapsto\left({\cal X}_{L_{S}},2^{{\cal X}_{L_{S}}}\right)$
is defined as the bijective and measurable function which uniquely
maps the $l$-th cell to the respective \textit{reconstruction point}
$\boldsymbol{x}_{L_{S}}^{l}$, $\forall l\in\mathbb{N}_{L_{S}}^{+}$,
according to some predefined ordering. That is, ${\cal Q}_{L_{S}}\left(\boldsymbol{x}\right)\triangleq\boldsymbol{x}_{L_{S}}^{l}$
if and only if $\boldsymbol{x}$ belongs to the respective cell (for
a detailed and more formalistic definition, see \cite{KalPetGRID2014}).
Having defined the quantizer ${\cal Q}_{L_{S}}\left(\cdot\right)$,
we consider the following discrete state space approximations of the
process $X_{t}$ \cite{Pages2005optimal}:
\begin{itemize}
\item The \textit{Markovian Quantization} of the state, defined as
\begin{equation}
\widetilde{X}_{t}^{L_{S}}\triangleq{\cal Q}_{L_{S}}\left(f\left(\widetilde{X}_{t-1}^{L_{S}},W_{t}\right)\right)\in{\cal X}_{L_{S}},\quad\forall t\in\mathbb{N}.
\end{equation}
where we have assumed explicitly apriori knowledge of a transition
mapping, modeling the temporal evolution of the Markov process $X_{t}$,
and
\item The \textit{Marginal Quantization} of the state, defined as
\begin{equation}
\overline{X}_{t}^{L_{S}}\triangleq{\cal Q}_{L_{S}}\left(X_{t}\right)\in{\cal X}_{L_{S}},\quad\forall t\in\mathbb{N}.
\end{equation}

\end{itemize}
Additionally, for later reference, define the column stochastic matrices
$\widetilde{\boldsymbol{P}}\in\left[0,1\right]^{L_{S}\times L_{S}}$
and $\overline{\boldsymbol{P}}\in\left[0,1\right]^{L_{S}\times L_{S}}$
as
\begin{flalign}
\widetilde{\boldsymbol{P}}\left(i,j\right) & \triangleq{\cal P}\left(\left.\widetilde{X}_{t}^{L_{S}}\equiv\boldsymbol{x}_{L_{S}}^{i}\right|\widetilde{X}_{t-1}^{L_{S}}\equiv\boldsymbol{x}_{L_{S}}^{j}\right)\quad\text{and}\\
\overline{\boldsymbol{P}}\left(i,j\right) & \triangleq{\cal P}\left(\left.\overline{X}_{t}^{L_{S}}\equiv\boldsymbol{x}_{L_{S}}^{i}\right|\overline{X}_{t-1}^{L_{S}}\equiv\boldsymbol{x}_{L_{S}}^{j}\right),
\end{flalign}
$\forall\left(i,j\right)\in\mathbb{N}_{L_{S}}^{+}\times\mathbb{N}_{L_{S}}^{+}$,
obviously related to the Markovian and marginal state quantizations,
respectively. Due to its structure, $\widetilde{\boldsymbol{P}}$
can at least be constructed simulating $\widetilde{X}_{t}^{L_{S}}$.
From the Law of Large Numbers, the entries of $\widetilde{\boldsymbol{P}}$
can be estimated with arbitrary precision from a sufficiently large
number of realizations of $\widetilde{X}_{t}^{L_{S}},t\in\mathbb{N}_{T}$,
for some $T<\infty$. Similarly, $\overline{\boldsymbol{P}}$ can
be estimated also with arbitrary precision from multiple realizations
of $\overline{X}_{t}^{L_{S}}$, which constitute a deterministic functional
of the true state $X_{t}$. Note, however, that in this case, it is
possible to obtain $\overline{\boldsymbol{P}}$ only using available
realizations of the state, \textit{without actually knowing either
the stochastic kernel or the transition mapping of }$X_{t}$ (if such
exists). For example, this could be made possible in sufficiently
controlled physical experiments, specially designed for system identification,
where the state $X_{t}$ wou\setlength{\textfloatsep}{5pt} ld be
a fully observable stochastic process.

\subsection{Asymptotically Optimal Recursive Estimators}

First, let us introduce the concept of \textit{conditional regularity}
of stochastic kernels and stochastic processes, which, as we will
see below, plays an important role in the asymptotic consistency of
a special class of approximate state estimators, based on the marginal
state quantization discussed above.
\begin{defn}
\label{Cond_Reg}\textbf{(Conditional Regularity of Stochastic Kernels
\cite{KalPetGRID2014})} Consider the stochastic (or Markov) kernel
${\cal K}:\mathscr{B}\left(\mathbb{R}^{M\times1}\right)\times\mathbb{R}^{M\times1}\mapsto\left[0,1\right]$,
associated with the process $Y_{t}\left(\omega\right)\in\mathbb{R}^{M\times1}$,
for all $t\in\mathbb{N}$. We say that ${\cal K}\left(\left.\cdot\right|\cdot\right)$
is \textit{Conditionally Regular of Type I (CRT I)}, if, for almost
all $\boldsymbol{x}\in{\cal Z}$, there exists a bounded sequence
$\delta_{n}\left(\boldsymbol{x}\right)\in\mathbb{R}_{+},n\in\mathbb{N}^{+}$
such that, for ${\cal A}\in\left\{ {\cal Z}_{L_{S}}^{j}\right\} _{j\in\mathbb{N}_{L_{S}}^{+}}$,\renewcommand{\arraystretch}{1.3}
\begin{equation}
\begin{array}{c}
{\displaystyle \sup_{{\cal A}}}\left|{\cal K}\left(\left.{\cal A}\right|\boldsymbol{x}\right)-{\cal K}\left({\cal A}\left|{\cal Q}_{L_{S}}\left(\boldsymbol{x}\right)\right.\right)\right|\le\dfrac{\delta_{L_{S}}\left(\boldsymbol{x}\right)}{L_{S}}\\
\text{with}\quad{\displaystyle \lim_{L_{S}\rightarrow\infty}}\delta_{L_{S}}\left(\boldsymbol{x}\right)\equiv0,\quad a.e.
\end{array},
\end{equation}
\renewcommand{\arraystretch}{1}If, additionally, for almost all $\boldsymbol{x}\in\mathbb{R}^{M\times1}$,
the Borel probability measure ${\cal K}\left(\left.\cdot\right|\boldsymbol{x}\right)$
admits a stochastic kernel density $\kappa:\mathbb{R}^{M\times1}\times\mathbb{R}^{M\times1}\mapsto\left[0,1\right],$
suggestively denoted as $\kappa\left(\left.\boldsymbol{y}\right|\boldsymbol{x}\right)$
and if the condition\renewcommand{\arraystretch}{1.3}
\begin{equation}
\begin{array}{c}
{\displaystyle \sup_{\boldsymbol{y}\in\mathbb{R}^{M\times1}}}\left|\kappa\left(\left.\boldsymbol{y}\right|\boldsymbol{x}\right)-\kappa\left(\left.\boldsymbol{y}\right|{\cal Q}_{L_{S}}\left(\boldsymbol{x}\right)\right)\right|\le\delta_{L_{S}}\left(\boldsymbol{x}\right)\\
\text{with}\quad{\displaystyle \lim_{L_{S}\rightarrow\infty}}\delta_{L_{S}}\left(\boldsymbol{x}\right)\equiv0,\quad a.e.
\end{array},
\end{equation}
\renewcommand{\arraystretch}{1}is satisfied, then we say that ${\cal K}\left(\left.\cdot\right|\cdot\right)$
is \textit{Conditionally Regular of Type II (CRT II)}. In any of the
two cases, we will also say that $Y_{t}$ is conditionally regular,
interchangeably.
\end{defn}
\vspace{-0.4cm}

\noindent \begin{center}
\rule[0.5ex]{0.5\columnwidth}{0.5pt}
\par\end{center}

\vspace{-0.2cm}

We are now ready to present the following two central results, establishing
that, under Assumption 1 and certain but mild assumptions on the nature
of the state process $X_{t}$, it is possible to approximate the strictly
optimal nonlinear filter/predictor $\widehat{X}_{t+\rho}$ by a simple
\textit{recursive} filtering scheme, being \textit{formally similar}
to the MMSE optimal filter of a Markov chain with finite state space.
The resulting approximate filter is strongly theoretically consistent,
in the sense that it converges to the true optimal filter of the state
uniformly inside each fixed finite time interval and uniformly in
a measurable set consisting of possible outcomes occurring with probability
almost $1$.

The results are presented below. The proofs are omitted, since each
one of them essentially constitutes a fusion of several related results
recently presented by the authors in \cite{KalPetGRID2014}. In the
following, ${\cal Q}_{L_{S}}^{e}:\left({\cal X}_{L_{S}},2^{{\cal X}_{L_{S}}}\right)\mapsto\left({\cal B}_{L_{S}},2^{{\cal B}_{L_{S}}}\right)$
constitutes a unique bijective mapping between the sets ${\cal X}_{L_{S}}$
and ${\cal B}_{L_{S}}\triangleq\left\{ {\bf e}_{l}^{L_{S}}\right\} _{l\in\mathbb{N}_{L_{S}}^{+}}$,
where the latter contains as elements the complete standard basis
in $\mathbb{R}^{L_{S}\times1}$.
\begin{thm}
\label{OUR_Filter}\textbf{\textup{(Approximate Filtering of Markov
Processes \cite{KalPetGRID2014})}} Define the \textbf{reconstruction}
and \textbf{likelihood} matrices as
\begin{flalign}
{\bf X} & \triangleq\left[\boldsymbol{x}_{L_{S}}^{1}\,\boldsymbol{x}_{L_{S}}^{2}\,\ldots\,\boldsymbol{x}_{L_{S}}^{L_{S}}\right]\in\mathbb{R}^{M\times L_{S}}\quad\text{and}\\
\boldsymbol{\Lambda}_{t} & \triangleq\mathrm{diag}\left(\left\{ \lambda_{t}\left(\boldsymbol{x}_{L_{S}}^{j}\right)\right\} _{j\in\mathbb{N}_{L_{S}}^{+}}\right)\in\mathbb{R}^{L_{S}\times L_{S}},
\end{flalign}
respectively, where, for all $t\in\mathbb{N}$, $\lambda_{t}\left(\boldsymbol{x}_{L_{S}}^{j}\right)$
is given by \eqref{eq:LR_SPECIFIC-1} (top of next page). Then, the
strictly optimal filter and $\rho$-step predictor of the state process
$X_{t}$ can be approximated as
\begin{equation}
{\cal E}^{L_{S}}\left(\left.X_{t+\rho}\right|\mathscr{Y}_{t}\right)\triangleq\dfrac{{\bf X}\boldsymbol{P}^{\rho}E_{t}}{\left\Vert E_{t}\right\Vert _{1}},\quad\forall t\in\mathbb{N}\label{eq:FILTER_1}
\end{equation}
and for all finite prediction horizons $\rho\ge0$, where the process
$E_{t}\in\mathbb{R}^{L_{S}\times1}$ on the RHS of \eqref{eq:FILTER_1}
satisfies the simple linear recursion
\begin{equation}
E_{t}\equiv\boldsymbol{\Lambda}_{t}\boldsymbol{P}E_{t-1},\quad\forall t\in\mathbb{N}
\end{equation}
and where 
\begin{itemize}
\item $\boldsymbol{P}\equiv\widetilde{\boldsymbol{P}}$, for the Markovian
quantization of the state and
\item $\boldsymbol{P}\equiv\overline{\boldsymbol{P}}$, for the marginal
quantization of the state.
\end{itemize}
The approximate filter ${\cal E}^{L_{S}}\left(\left.X_{t}\right|\mathscr{Y}_{t}\right)$
is initialized at time $t\equiv-1$ setting $E_{-1}\equiv\mathbb{E}\left\{ {\cal Q}_{L_{S}}^{e}\left(\widetilde{X}_{-1}^{L_{S}}\right)\right\} $
for the Markovian quantization and $E_{-1}\equiv\mathbb{E}\left\{ {\cal Q}_{L_{S}}^{e}\left(\overline{X}_{-1}^{L_{S}}\right)\right\} $
for the marginal quantization of the state.
\end{thm}

\begin{figure*}[t!]
\hrulefill
\normalsize
\begin{equation}
\lambda_{t}\left(\boldsymbol{x}_{L_{S}}^{j}\right)\triangleq\dfrac{\exp\left(-\dfrac{1}{2}\left({\bf y}_{t}-\boldsymbol{A}_{t}\boldsymbol{x}_{L_{S}}^{j}\right)^{\boldsymbol{T}}\left(\boldsymbol{\Sigma}_{t}\left(\boldsymbol{x}_{L_{S}}^{j}\right)+\sigma_{\xi}^{2}{\bf I}_{N\times N}\right)^{-1}\left({\bf y}_{t}-\boldsymbol{A}_{t}\boldsymbol{x}_{L_{S}}^{j}\right)\right)}{\sqrt{\det\left(\boldsymbol{\Sigma}_{t}\left(\boldsymbol{x}_{L_{S}}^{j}\right)+\sigma_{\xi}^{2}{\bf I}_{N\times N}\right)}},\label{eq:LR_SPECIFIC-1}
\end{equation}
\hrulefill
\end{figure*}
\begin{thm}
\label{OUR_Filter-1}\textbf{\textup{(Asymptotic Optimality of Approximate
Filters \cite{KalPetGRID2014})}} Pick any natural $T<\infty$ and
suppose either of the following:
\begin{itemize}
\item The Markovian quantization is employed, whose initial value coincides
with that of $X_{t}$, and the transition mapping of the state, $f:{\cal Z}\times{\cal W}\overset{a.s.}{\mapsto}{\cal Z}$,
is Lipschitz in ${\cal Z}$, for every element of ${\cal W}$.
\item The marginal quantization is employed and $X_{t}$ is conditionally
regular.
\end{itemize}
Then, for any finite prediction horizon $\rho\ge0$, there exists
a measurable subset $\widehat{\Omega}_{T}\subseteq\Omega$ with ${\cal P}$-measure
at least $1-\left(T+1\right)^{1-CN}\exp\left(-CN\right)$, such that
\begin{equation}
\sup_{t\in\mathbb{N}_{T}}\sup_{\omega\in\widehat{\Omega}_{T}}\left\Vert {\cal E}^{L_{S}}\left(\left.X_{t+\rho}\right|\mathscr{Y}_{t}\right)-\widehat{X}_{t+\rho}\right\Vert _{1}\underset{L_{S}\rightarrow\infty}{\longrightarrow}0,
\end{equation}
for any free, finite constant $C\ge1$. In other words, the convergence
of the respective approximate filters is compact in $t\in\mathbb{N}$
and, with probability at least $1-\left(T+1\right)^{1-CN}\exp\left(-CN\right)$,
uniform in $\omega$.
\end{thm}
Note that while the approximate filters/predictors described in Theorem
\ref{OUR_Filter} are structurally very simple, they converge to the
respective optimal nonlinear state estimators in a particularly strong
sense (under the respective conditions), as Theorem \ref{OUR_Filter-1}
clearly suggests.

\section{SCST \& SSCP in Mobile Wireless Networks}

In this section, we present the main results of the paper. In a nutshell,
we propose two theoretically consistent sequential algorithms for
approximately solving the SCST and SSCP problems defined in Section
II.B, both derived as applications of Theorem \ref{OUR_Filter}, presented
in Section III.

\subsection{SCST}

At this point, it is apparent that Theorem \ref{OUR_Filter} in fact
directly provides us with an effective approximate and recursive estimator
for the channel state $X_{t}$. Therefore, Theorem \ref{OUR_Filter}
immediately solves the SCST problem, since the resulting filtering/prediction
scheme is sequential and, as new channel measurements become available,
its computational complexity is fixed, due to time invariance of the
type of numerical operations required for each filter update.

Specifically, Algorithm 1 shows the discrete steps required for the
centralized implementation of the proposed filtering scheme, in a
relatively powerful fusion center. Observe that, depending on the
type of quantization employed and for each $\rho\ge0$, the required
matrices $\boldsymbol{P}$ and ${\bf X}\boldsymbol{P}^{\rho}$ can
be computed offline and stored in memory.

\begin{algorithm}
\begin{enumerate}
\item Choose $\boldsymbol{P}$ and $E_{-1}$ depending on the type of state
quantization employed (Markovian or marginal).
\item Choose $\rho\ge0$ and recall $\boldsymbol{P}$ and ${\bf X}\boldsymbol{P}^{\rho}$
from memory.
\item \textbf{For $t=0,1,...$ do}
\item \quad{}Compute the diagonal matrix $\boldsymbol{\Lambda}_{t}$ from
\eqref{eq:LR_SPECIFIC-1}.
\item \quad{}Compute \& store 
\[
E_{t}=\boldsymbol{\Lambda}_{t}\boldsymbol{P}E_{t-1},
\]
\quad{}until the next iteration.
\item \quad{}Normalize $E_{t}$ as
\[
E_{t}^{N}=\dfrac{E_{t}}{\left\Vert E_{t}\right\Vert _{1}}.
\]

\item \quad{}Compute \& output $E_{t}$ as
\[
{\cal E}^{L_{S}}\left(\left.X_{t+\rho}\right|\mathscr{Y}_{t}\right)={\bf X}\boldsymbol{P}^{\rho}E_{t}^{N}.
\]

\item \textbf{endFor}
\end{enumerate}
\protect\caption{Sequential Channel State Tracking}

\end{algorithm}

\subsection{SSCP}

Defining the natural filtration generated by both the state $X_{t}$
and the observations ${\bf y}_{t}$ as
\begin{equation}
\left\{ \mathscr{H}_{t}\right\} _{t\in\mathbb{N}}\triangleq\left\{ \sigma\left\{ \left\{ X_{t},{\bf y}_{i}\right\} _{i\in\mathbb{N}_{t}}\right\} \right\} _{t\in\mathbb{N}}
\end{equation}
and using the tower property of expectations, it is true that
\begin{align}
\widehat{y}_{t}\left({\bf q}\right) & \equiv\mathbb{E}\left\{ \left.y_{t}\left({\bf q}\right)\right|\mathscr{Y}_{t}\right\} \nonumber \\
 & \equiv\mathbb{E}\left\{ \left.\mathbb{E}\left\{ \left.y_{t}\left({\bf q}\right)\right|\mathscr{H}_{t}\right\} \right|\mathscr{Y}_{t}\right\} ,\quad\forall t\in\mathbb{N}.\label{eq:Spatial_1}
\end{align}
Let us define the quantities
\begin{flalign}
\alpha^{{\bf q}} & \triangleq-10\log_{10}\left(\left\Vert {\bf q}-{\bf p}_{ref}\right\Vert _{2}\right),\\
\left.\sigma^{{\bf q}}\left(X_{t}\right)\right|X_{t} & \overset{def}{\sim}{\cal N}\left(0,\eta^{2}\left(X_{t}\right)\right)\quad\text{and}\\
\xi^{{\bf q}} & \overset{def}{\sim}{\cal N}\left(0,\sigma_{\xi}^{2}\right),
\end{flalign}
where, also by definition,
\begin{align}
\mathbb{E}\left\{ \left.\begin{bmatrix}\boldsymbol{\sigma}_{t}\left(X_{t}\right)\\
\sigma^{{\bf q}}\left(X_{t}\right)
\end{bmatrix}\begin{bmatrix}\boldsymbol{\sigma}_{t}\left(X_{t}\right)\\
\sigma^{{\bf q}}\left(X_{t}\right)
\end{bmatrix}^{\boldsymbol{T}}\right|X_{t}\right\}  & \triangleq\begin{bmatrix}\boldsymbol{\Sigma}_{t}\left(X_{t}\right) & \boldsymbol{\sigma}_{t}^{{\bf q}}\left(X_{t}\right)\\
\left(\boldsymbol{\sigma}_{t}^{{\bf q}}\left(X_{t}\right)\right)^{\boldsymbol{T}} & \eta^{2}\left(X_{t}\right)
\end{bmatrix},
\end{align}
with each element of $\boldsymbol{\sigma}_{t}^{{\bf q}}\left(X_{t}\right)\in\mathbb{R}^{N\times1}$
given by
\begin{equation}
\boldsymbol{\sigma}_{t}^{{\bf q}}\left(X_{t}\right)\left(j\right)\triangleq{\cal {\cal R}}\left({\bf q},{\bf p}_{j}\left(t\right),\boldsymbol{\theta}\left(X_{t}\right)\right),\quad\forall j\in\mathbb{N}_{N}^{+}.
\end{equation}
Then, it must be true that
\begin{flalign}
y_{t}\left({\bf q}\right) & \equiv\alpha^{{\bf q}}X_{t}\left(1\right)+\sigma^{{\bf q}}\left(X_{t}\right)+\xi^{{\bf q}}\nonumber \\
 & \equiv A^{{\bf q}}X_{t}+\sigma^{{\bf q}}\left(X_{t}\right)+\xi^{{\bf q}},
\end{flalign}
where $A^{{\bf q}}\triangleq\left[\alpha^{{\bf q}}\,{\bf 0}_{1\times\left(M-1\right)}\right]\in\mathbb{R}^{1\times M}$,
since $y_{t}\left({\bf q}\right)$ can be equivalently considered
as an additional observation, measured by an imaginary sensor at position
${\bf q}$, which of course was not used for state estimation in the
SCST problem treated above. Under these considerations, the inner
conditional expectation of \eqref{eq:Spatial_1} can be expressed
as
\begin{flalign}
\mathbb{E}\left\{ \left.y_{t}\left({\bf q}\right)\right|\mathscr{H}_{t}\right\}  & \equiv\mathbb{E}\left\{ \left.A^{{\bf q}}X_{t}+\sigma^{{\bf q}}\left(X_{t}\right)+\xi^{{\bf q}}\right|\mathscr{H}_{t}\right\} \nonumber \\
 & \equiv A^{{\bf q}}X_{t}+\mathbb{E}\left\{ \left.\sigma^{{\bf q}}\left(X_{t}\right)\right|\mathscr{H}_{t}\right\} ,
\end{flalign}
or, using well known properties of jointly Gaussian random vectors
\cite{Papoulis2002} (also used in \cite{MostofiSpatial2012}),
\begin{align}
\mathbb{E}\left\{ \left.y_{t}\left({\bf q}\right)\right|\mathscr{H}_{t}\right\}  & =A^{{\bf q}}X_{t}+\left(\boldsymbol{\sigma}_{t}^{{\bf q}}\left(X_{t}\right)\right)^{\boldsymbol{T}}{\bf C}_{t}^{-1}\left(X_{t}\right)\left({\bf y}_{t}-\boldsymbol{A}_{t}X_{t}\right)\triangleq\phi_{t}\left(X_{t},{\bf y}_{t}\right).\label{eq:Spatial_3}
\end{align}
As a result, $\widehat{y}_{t}\left({\bf q}\right)$ can be expressed
as
\begin{equation}
\widehat{y}_{t}\left({\bf q}\right)\equiv\mathbb{E}\left\{ \left.\phi_{t}\left(X_{t},{\bf y}_{t}\right)\right|\mathscr{Y}_{t}\right\} ,\label{eq:Spatial_4}
\end{equation}
that is, the SSCP problem coincides with the problem of sequentially
evaluating the optimal nonlinear filter of a particular functional
of the state and the observations, $\phi_{t}\left(\cdot,\cdot\right)$.
In this respect, the following result is true, which, together with
the results presented in Section III, has also been formulated previously
by the authors in \cite{KalPetGRID2014}.

\begin{thm}
\label{Functionals}\textbf{\textup{(Approximate Filtering for Functionals
of the State / Separation Theorem \cite{KalPetGRID2014})}} For any
deterministic functional family $\left\{ \boldsymbol{\phi}_{t}:\mathbb{R}^{L_{S}\times1}\mapsto\mathbb{R}^{M_{\phi_{t}}\times1}\right\} _{t\in\mathbb{N}}$
with bounded and continuous members and any finite prediction horizon
$\rho\ge0$, the strictly optimal filter and $\rho$-step predictor
of the \textbf{transformed} \textbf{process} $\boldsymbol{\phi}_{t}\left(X_{t}\right)$
can be approximated as
\begin{equation}
{\cal E}^{L_{S}}\left(\left.\boldsymbol{\phi}_{t+\rho}\left(X_{t+\rho}\right)\right|\mathscr{Y}_{t}\right)\triangleq\boldsymbol{\Phi}_{t+\rho}\dfrac{\boldsymbol{P}^{\rho}E_{t}}{\left\Vert E_{t}\right\Vert _{1}}\in\mathbb{R}^{M_{\phi_{t}}\times1},\label{eq:Spatial_2}
\end{equation}
for all $t\in\mathbb{N}$, where the process $E_{t}\in\mathbb{R}^{L_{S}\times1}$
can be recursively evaluated as in Theorem \ref{OUR_Filter} and 
\begin{equation}
\boldsymbol{\Phi}_{t+\rho}\triangleq\left[\boldsymbol{\phi}_{t+\rho}\left(\boldsymbol{x}_{L_{S}}^{1}\right)\,\ldots\,\boldsymbol{\phi}_{t+\rho}\left(\boldsymbol{x}_{L_{S}}^{L_{S}}\right)\right]\in\mathbb{R}^{M_{\phi_{t}}\times L_{S}}.
\end{equation}
In the above, the transition matrix $\boldsymbol{P}$ and the initialization
of the approximate filter are exactly the same as in Theorem \ref{OUR_Filter}.
Additionally, the approximate filter is asymptotically optimal under
the same conditions and in the same sense as in Theorem \ref{OUR_Filter-1}.
\end{thm}
Invoking Theorem \ref{Functionals}, the following result is true,
providing a closed form approximate solution to the SSCP problem,
at the same time enjoying asymptotic optimality in the sense of Theorem
\ref{OUR_Filter-1}.
\begin{thm}
\label{SSCP}\textbf{\textup{(Approximate Solution to the SSCP Problem)}}
The strictly optimal spatiotemporal predictor of the channel magnitude
at an arbitrary position ${\bf q}\in\mathbb{R}^{2}$, $\widehat{y}_{t}\left({\bf q}\right)$,
can be approximated as\makeatletter
\renewcommand*\env@cases[1][2]{
\let\@ifnextchar\new@ifnextchar
\left\lbrace   
\def\arraystretch{#1}%
\array{@{}l@{\quad}l@{}}%
}
\makeatother 
\begin{equation}
{\cal E}^{L_{S}}\left(\left.y_{t+\rho}\left({\bf q}\right)\right|\mathscr{Y}_{t}\right)\triangleq\begin{cases}
{\displaystyle \left\langle \boldsymbol{\phi}_{t}\left({\bf y}_{t}\right),\dfrac{E_{t}}{\left\Vert E_{t}\right\Vert _{1}}\right\rangle }, & \rho\equiv0\\
A^{{\bf q}}{\bf X}\boldsymbol{P}^{\rho}\dfrac{E_{t}}{\left\Vert E_{t}\right\Vert _{1}}, & \rho\ge1
\end{cases},\label{eq:Spatial_2-1}
\end{equation}
for all $t\in\mathbb{N}$, where the process $E_{t}\in\mathbb{R}^{L_{S}\times1}$
can be recursively evaluated as in Theorem \ref{OUR_Filter} and where
the stochastic process $\boldsymbol{\phi}_{t}\left({\bf y}_{t}\right)\in\mathbb{R}^{L_{S}\times1}$
is defined as 
\begin{equation}
\boldsymbol{\phi}_{t}\left({\bf y}_{t}\right)\triangleq\left[\phi_{t}\left(\boldsymbol{x}_{L_{S}}^{1},{\bf y}_{t}\right)\,\ldots\,\phi_{t}\left(\boldsymbol{x}_{L_{S}}^{L_{S}},{\bf y}_{t}\right)\right]^{\boldsymbol{T}},\label{eq:Spatial_6}
\end{equation}
with $\phi_{t}:\mathbb{R}^{M\times1}\times\mathbb{R}^{N\times1}\mapsto\mathbb{R}$
defined as in \eqref{eq:Spatial_3}. In the above, the transition
matrix $\boldsymbol{P}$ and the initialization of the approximate
filter are exactly the same as in Theorem \ref{OUR_Filter}. Additionally,
under the same conditions as in Theorem \ref{OUR_Filter-1}, it is
true that
\begin{equation}
\sup_{t\in\mathbb{N}_{T}}\sup_{\omega\in\widehat{\Omega}_{T}}\left|{\cal E}^{L_{S}}\left(\left.y_{t+\rho}\left({\bf q}\right)\right|\mathscr{Y}_{t}\right)-\widehat{y}_{t+\rho}\left({\bf q}\right)\right|\underset{L_{S}\rightarrow\infty}{\longrightarrow}0.
\end{equation}

\end{thm}
\begin{algorithm}
\begin{enumerate}
\item Choose $\boldsymbol{P}$ and $E_{-1}$ depending on the type of state
quantization employed (Markovian or marginal).
\item Choose $\rho\ge0$ and recall $\boldsymbol{P}$ and ${\bf X}\boldsymbol{P}^{\rho}$
from memory.
\item Choose an arbitrary ${\bf q}\in\mathbb{R}^{2}$.
\item \textbf{For $t=0,1,...$ do}
\item \quad{}Compute the diagonal matrix $\boldsymbol{\Lambda}_{t}$ from
\eqref{eq:LR_SPECIFIC-1}.
\item \quad{}Compute 
\[
E_{t}=\boldsymbol{\Lambda}_{t}\boldsymbol{P}E_{t-1},
\]
\quad{}\& store it until the next iteration.
\item \quad{}Normalize $E_{t}$ as
\[
E_{t}^{N}=\dfrac{E_{t}}{\left\Vert E_{t}\right\Vert _{1}}.
\]

\item \quad{}Compute vector $\boldsymbol{\phi}_{t}\left({\bf y}_{t}\right)$
using current observations\linebreak{}
\hphantom{\quad{}}from \eqref{eq:Spatial_3} and \eqref{eq:Spatial_6}.
\item \quad{}Compute \& output ${\cal E}^{L_{S}}\left(\left.y_{t+\rho}\left({\bf q}\right)\right|\mathscr{Y}_{t}\right)$
as\makeatletter
\renewcommand*\env@cases[1][2]{
\let\@ifnextchar\new@ifnextchar
\left\lbrace   
\def\arraystretch{#1}%
\array{@{}l@{\quad}l@{}}%
}
\makeatother
\[
{\cal E}^{L_{S}}\left(\left.y_{t+\rho}\left({\bf q}\right)\right|\mathscr{Y}_{t}\right)\triangleq\begin{cases}
{\displaystyle \left\langle \boldsymbol{\phi}_{t}\left({\bf y}_{t}\right),E_{t}^{N}\right\rangle }, & \rho\equiv0\\
A^{{\bf q}}{\bf X}\boldsymbol{P}^{\rho}E_{t}^{N}, & \rho\ge1
\end{cases}.
\]

\item \textbf{endFor}
\end{enumerate}
\protect\caption{Sequential Spatiotemporal Channel Prediction}
\end{algorithm}

\begin{IEEEproof}[Proof of Theorem \ref{SSCP}]
See Appendix.
\end{IEEEproof}
Algorithm 2 shows the discrete steps required for the centralized
implementation of the proposed approximate spatial prediction scheme.

\subsection{Computational Complexity: A Fair Comparison}

A careful inspection of Algorithms 1 and 2 proposed earlier for the
solution of the SCST and SSCP problems, respectively, reveals that,
in the worst case, the computational complexity of both algorithms
scales as ${\cal O}\left(L_{S}^{2}+L_{S}N^{3}\right)$. The two algorithms
can also be combined into one with the same computational requirements.
The cubic term related to the number of sensors in the network is
due to the inversion and the determinant calculation of the covariance
matrices, for each reconstruction point $\boldsymbol{x}_{L_{S}}^{j},j\in\mathbb{N}_{L_{S}}^{+}$
and at each time instant $t\in\mathbb{N}$, and it is computationally
bearable, at least for a relatively small number of sensors. However,
note that when the sensors are stationary or even when their trajectories
are fixed and known apriori, the aforementioned computationally demanding
operations may be completely bypassed by precomputing the required
matrices for each set of parameters and storing them in memory. In
such a case, the computational complexity of both algorithms reduces
significantly to ${\cal O}\left(L_{S}^{2}+L_{S}N^{2}\right)$.

Temporarily considering the number of sensors $N$ as constant and
focusing solely on the number of quantization regions $L_{S}$, it
is apparent that the complexity of both algorithms considered scales
as ${\cal O}\left(L_{S}^{2}\right)$. This can be very large if one
considers the same quantization resolution in each dimension of the
Euclidean space the state process lives in, that is, as in Section
III.A, where, for simplicity, we considered a completely uniform strictly
hypercubic quantizer on the set $\left[a,b\right]^{M}$. Specifically,
in this general case, where we ``pay the same attention'' to all
points in the $M$-hypercube of interest, the overall complexity scales
as ${\cal O}\left(L_{S}^{2}\equiv L^{2M}\right)$, which, of course,
is clearly prohibitively large for high dimensional systems. However,
as analyzed in \cite{KalPetGRID2014}, no one prevents us from considering
either hyperrectangular Euclidean state spaces since, in most cases,
each element of the state vector would have its own dynamic range,
or different quantization resolution for each element, since they
may not have all the same importance in the particular engineering
application of interest, or both.

In the particular problems we are interested in here, the dimension
of the state process is relatively low, that is, the range of $M$
almost always between 2 and 5 dimensions, which makes grid based filters
practically feasible. Additionally, as we will see in Section V, where
we present the relevant numerical simulations and as it has already
been clearly shown in \cite{MostofiSpatial2012} in a non Bayesian
framework, if one considers the simple isotropic autocorrelation kernel
described by \eqref{eq:isotropic} in Section II.B and is primarily
interested in the SSCP problem, the sensitivity of the quality of
spatial channel prediction on the estimation error of the shadowing
power and decorrelation distance of the channel is indeed very low,
making it possible to potentially consider a lower quantization resolution
for the aforementioned quantities without significant compromise in
terms of the prediction quality. As a result, grid based approximate
filters are indeed adequate for the problems of interest in this paper,
taking advantage of their strong properties in terms of asymptotic
consistency.

Naturally, particle filters \cite{PARTICLE2002tutorial,Gordon1993Novel}
would constitute the rivals of our grid based filtering approach.
Compared to the former, particle filters exhibit a computational complexity
of ${\cal O}\left(L_{S}\right)$, where, in this case, $L_{S}$ constitutes
the number of particles \cite{PARTICLE2002tutorial}. That is, the
complexity of particle filters is one order of magnitude smaller compared
to the complexity of grid based filters. Note, though, that in Algorithms
1 and 2, the one and only computational operation which incurs a complexity
of ${\cal O}\left(L_{S}^{2}\right)$ is the matrix vector operation
$\boldsymbol{P}E_{t-1}$. In fact, in the numerical simulations conducted
in Section V, it was revealed that, at least for the problems of interest,
the numerical operations of inversion and determinant computation
of the respective covariance matrices are far more computationally
important than the aforementioned matrix vector multiplication. These
operations would be also required in a typical particle filter implementation
as well.

Continuing the comparison of the proposed approach with the filtering
approximation family of particle filters, another issue of major importance
is filter behavior with respect to the curse of dimensionality. Although
in, for instance, \cite{Crisan2002Survey} and \cite{VikramDoucet2001Particle},
it was warmly asserted that the use of particle filters might indeed
make it possible to beat the curse of dimensionality, it was later
made clear that this is not the case and that particle filters indeed
suffer from the curse \cite{DAUM_CURSE2003,Bengtsson2008Curse,Quang2010Insight}.
More recently, it was shown that particle filters suffer in general
both in terms of temporal uniformity in the convergence of the respective
filtering approximations and in terms of exponential dependence on
the dimensionality of the observation process \cite{RebeschiniRamon2013_1}
and that of the state process as well \cite{Quang2010Insight}, greatly
affecting rate of convergence. In fact, as also shown in \cite{RebeschiniRamon2013_1},
in order to somewhat circumvent these important practical limitations,
strict assumptions must hold regarding the structure of the partially
observable system under consideration, therefore somewhat limiting
the general applicability of the respective methods presented in \cite{RebeschiniRamon2013_1}.
Of course, the grid-based filters proposed in this paper suffer from
similar drawbacks \cite{KalPetGRID2014}, a fact that strengthens
the common belief that, at least in the context of nonlinear filtering,
the curse of dimensionality constitutes a ubiquitous phenomenon.

However, from a technical point of view, the grid based filters we
propose for effectively solving the SCST and SSCP problems are very
consistent, in the sense that their convergence to the true nonlinear
filter of the state is compact in time and uniform with respect to
a set of possible outcomes of almost full probability measure. Although,
due to our inability to show uniform convergence in time (however
keeping the class of admissible hidden Markov process large), we cannot
theoretically prove that the proposed approximate filters can indeed
reach a stable steady state, we can at least guarantee that the approximation
error will be uniformly bounded for any fixed time interval set by
the user, with overwhelmingly high probability (for more details,
see \cite{KalPetNonlinearEXTENDED2014} and \cite{KalPetGRID2014}).
Further, as we will see in the next section, the practical performance
of the filters, at least when considering the autocorrelation kernel
of \eqref{eq:isotropic}, is very robust and tracks the hidden system
accurately, for a relatively small number of quantization cells, without
the need of any fine tuning, as opposed to the case of particle filters
(choice of importance density, etc.) \cite{PARTICLE2002tutorial}.

\section{Numerical Simulations}

\begin{figure*}
\centering\subfloat[\label{fig:Channel_Tracking}]{\centering\includegraphics[clip,scale=0.6]{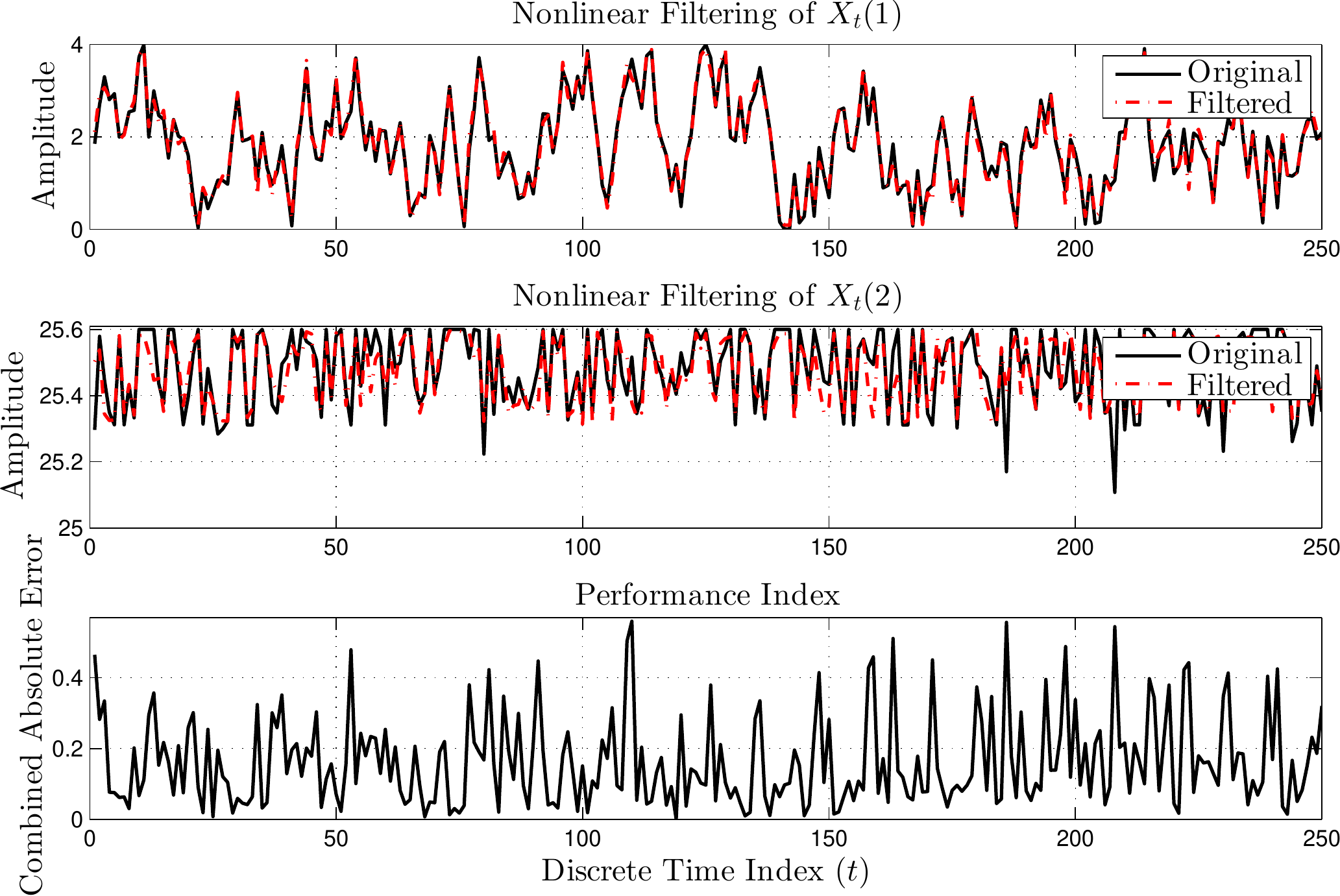}

}\\
\subfloat[\label{fig:Spatial_Pred}]{\centering\includegraphics[scale=0.6]{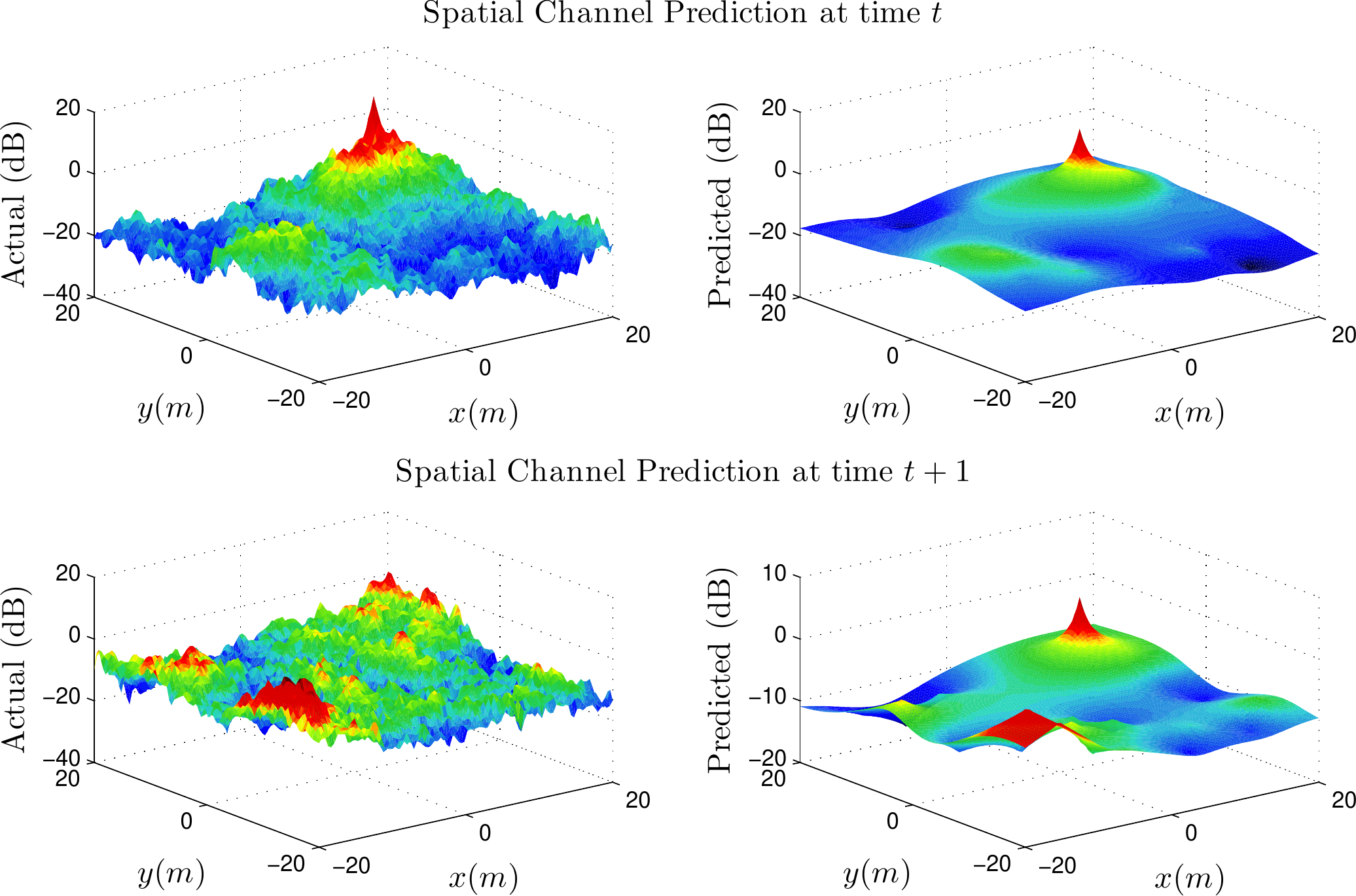}}\protect\caption{(a) Demonstration of channel state tracking for 250 time steps. The
estimates are produced from the observations of 30 randomly scattered
sensors in the square region $\left[-20\, m,20\, m\right]^{2}$. (b)
Spatial prediction and temporal tracking of the channel combined.
In this example, the spatial grid consists of 3600 evaluation points
and the results are obtained using just 30 spatial measurements. Observe
how the prediction procedure dynamically captures the basic characteristics
of the channel, exploiting the spatial correlations due to shadowing. }
\end{figure*}

The practical effectiveness of Algorithms 1 and 2 will also be validated
through a number of sufficiently representative synthetic experiments.
Specifically, we consider $N\equiv30$ sensors randomly scattered
on a sufficiently fine square grid, in the square region of the $x/y$
plane ${\cal S}\equiv\left[0,40\right]^{2}$ (in $m\times m$). The
position of the reference antenna is fixed at ${\bf p}_{ref}\equiv\left[25\,10\right]^{\boldsymbol{T}}$.
Concerning the behavior of the communication channel throughout the
plane, the variance of the multipath fading noise term is set at $\sigma_{\xi}^{2}\equiv2$
and, as far as shadowing is concerned, the autocorrelation kernel
of \eqref{eq:isotropic} is employed, where, for simplicity, we assume
that the correlation distance is known and constant with respect to
time and equal to $10\, m$. As a result, in this simple example,
the channel state is two dimensional, with $X_{t}\left(1\right)\equiv\mu\left(X_{t}\right)$
and $X_{t}\left(2\right)\equiv\boldsymbol{\theta}\left(X_{t}\right)$
representing the path loss coefficient and the shadowing power, respectively.
The temporal evolution of each respective component of the channel
state is given by the stochastic difference equations
\begin{flalign}
X_{t}\left(1\right) & \equiv\tanh\left(\gamma\left(X_{t-1}\left(1\right)-2\right)\right)+W_{t}+2,\quad\text{and}\\
X_{t}\left(2\right) & \equiv0.3\left|\tanh\left(\sin\left(\gamma X_{t-1}\left(2\right)W_{t}\right)\right.\right.+\nonumber \\
 & \hspace{3em}\left.\left.+X_{t-1}\left(2\right)W_{t}\right)+W_{t}\right|+25,\quad\forall t\in\mathbb{N},
\end{flalign}
for some arbitrary but known initial conditions, where $\gamma\equiv1.6$
and $W_{t}\equiv\text{clip}_{\left[-1,1\right]}\left(G_{t}\right)$,
$G_{t}\overset{i.i.d.}{\sim}{\cal N}\left(0,1\right)$, with $\text{clip}_{\left[-1,1\right]}\left(\cdot\right)$
denoting the hard limiter operation into the set $\left[-1,1\right]$.
Note that both difference equations are strongly nonlinear in in both
the state and driving noise. Also note the strong coupling between
the two equations, due to the fact that both are driven by exactly
the same noise realizations. The above equations attempt to model
a situation where the path loss exponent is somewhat slowly varying
between $0$ and $4$, whereas the shadowing power is rapidly varying
between $25$ and $25.6$. The state $X_{t}\equiv\left[X_{t}\left(1\right)\, X_{t}\left(2\right)\right]^{\boldsymbol{T}}$
was uniformly quantized into $L_{S}\equiv30^{2}$ cells (that is,
$L\equiv30$). For simplicity, regarding the simulation results which
will be presented and discussed below, we focus on the case where
$\rho\equiv0$, that is, we consider the problems of \textit{temporal
filtering of the channel state} and \textit{spatial prediction of
the channel}, which both constitute instances of the SCST and SSCP
problems, respectively.

Fig. \ref{fig:Channel_Tracking} demonstrates the channel state tracking
(temporal filtering of the state) for $250$ time steps, according
to the experimental setting stated above. As illustrated in the figure,
the quality of the estimates is very good, considering the nonlinearity
present both in the state process and the observations at each sensor
in the network. It also apparent that the produced estimation process
behaves in a stable manner, as time increases.

The filter of the channel state can subsequently be used for the also
asymptotically optimal prediction of the channel magnitude in the
rest of the space. This is illustrated in Fig. \ref{fig:Spatial_Pred},
where the combined spatial prediction and temporal tracking of the
channel magnitudes are shown (for two time instants), in comparison
to the real channel maps in the square region under consideration.
The random field used for modeling the spatial channel process was
generated using a spatial grid of $3600$ points and the respective
predicted values were obtained from just $N\equiv30$ random scattered
spatial channel measurements in the region of interest. From the figure,
it can be seen that the quality of the predicted process is very good,
especially considering the fact that the channel is reconstructed
using only $0.83\,\%$ of the total number of grid points in the region
of interest. Of course, the quality of the spatial prediction improves
as the number of spatial measurements (and therefore nodes/sensors)
increases. One can observe that the prediction procedure accurately
captures the basic characteristics of the channel in the region of
interest, effectively exploiting the spatial correlations due to shadowing.

\section{Conclusion}

In this paper, a nonlinear filtering framework was proposed for addressing
the fundamental problems of sequential channel state tracking and
spatiotemporal channel prediction in mobile wireless sensor networks.
First, we formulated the channel observations at each sensor as a
partially observable nonlinear system with temporally varying state
and spatiotemporally varying observations. Then, a grid based approximate
filtering scheme was employed for accurately tracking the temporal
variation of the channel state, based on which we proposed a recursive
spatiotemporal channel gain predictor, providing real time sequential
CG map estimation at each sensor in the network. Further, we showed
that both estimators are asymptotically optimal, in the sense that
they converge to the optimal MMSE estimators/predictors of the channel
state and observations at unobserved positions in the region of interest,
respectively, in a technically strong sense. In addition to these
theoretical results, numerical simulations were presented, validating
the practical effectiveness of the proposed approach and increasing
the user's confidence for practical consideration in real world wireless
networks.

\section*{Appendix\protect \\
Proof of Theorem \ref{SSCP}}

Let us first consider the filtering case, that is, the one where $\rho\equiv0$.
Substituting \eqref{eq:Spatial_3} to \eqref{eq:Spatial_4}, we get
\begin{multline}
\widehat{y}_{t}\left({\bf q}\right)\equiv\mathbb{E}\left\{ \left.A^{{\bf q}}X_{t}+\left(\boldsymbol{\sigma}_{t}^{{\bf q}}\left(X_{t}\right)\right)^{\boldsymbol{T}}{\bf C}_{t}^{-1}\left(X_{t}\right)\left({\bf y}_{t}-\boldsymbol{A}_{t}X_{t}\right)\right|\mathscr{Y}_{t}\right\} \\
\equiv A^{{\bf q}}\mathbb{E}\left\{ \left.X_{t}\right|\mathscr{Y}_{t}\right\} +\mathbb{E}\left\{ \left.\left(\boldsymbol{\sigma}_{t}^{{\bf q}}\left(X_{t}\right)\right)^{\boldsymbol{T}}{\bf C}_{t}^{-1}\left(X_{t}\right)\right|\mathscr{Y}_{t}\right\} {\bf y}_{t}-\\
-\mathbb{E}\left\{ \left.\left(\boldsymbol{\sigma}_{t}^{{\bf q}}\left(X_{t}\right)\right)^{\boldsymbol{T}}{\bf C}_{t}^{-1}\left(X_{t}\right)\boldsymbol{A}_{t}X_{t}\right|\mathscr{Y}_{t}\right\} ,\quad\forall t\in\mathbb{N},
\end{multline}
from where, defining the bounded and continuous functionals
\begin{flalign}
\boldsymbol{\phi}_{t}^{1}\left(X_{t}\right) & \triangleq\left(\left(\boldsymbol{\sigma}_{t}^{{\bf q}}\left(X_{t}\right)\right)^{\boldsymbol{T}}{\bf C}_{t}^{-1}\left(X_{t}\right)\right)^{\boldsymbol{T}}\in\mathbb{R}^{N\times1}\quad\text{and}\\
\phi_{t}^{2}\left(X_{t}\right) & \triangleq\left(\boldsymbol{\sigma}_{t}^{{\bf q}}\left(X_{t}\right)\right)^{\boldsymbol{T}}{\bf C}_{t}^{-1}\left(X_{t}\right)\boldsymbol{A}_{t}X_{t}\in\mathbb{R},
\end{flalign}
we can write
\begin{align}
\widehat{y}_{t}\left({\bf q}\right) & \equiv\mathbb{E}\left\{ \left.y_{t}\left({\bf q}\right)\right|\mathscr{Y}_{t}\right\} \equiv A^{{\bf q}}\mathbb{E}\left\{ \left.X_{t}\right|\mathscr{Y}_{t}\right\} +\left(\mathbb{E}\left\{ \left.\boldsymbol{\phi}_{t}^{1}\left(X_{t}\right)\right|\mathscr{Y}_{t}\right\} \right)^{\boldsymbol{T}}{\bf y}_{t}-\mathbb{E}\left\{ \left.\phi_{t}^{2}\left(X_{t}\right)\right|\mathscr{Y}_{t}\right\} .
\end{align}
Then, for all $t\in\mathbb{N}$, define the approximate operator
\begin{align}
{\cal E}^{L_{S}}\left(\left.y_{t}\left({\bf q}\right)\right|\mathscr{Y}_{t}\right) & \triangleq A^{{\bf q}}{\cal E}^{L_{S}}\left(\left.X_{t}\right|\mathscr{Y}_{t}\right)+\left({\cal E}^{L_{S}}\left(\left.\boldsymbol{\phi}_{t}^{1}\left(X_{t}\right)\right|\mathscr{Y}_{t}\right)\right)^{\boldsymbol{T}}{\bf y}_{t}-{\cal E}^{L_{S}}\left(\left.\phi_{t}^{2}\left(X_{t}\right)\right|\mathscr{Y}_{t}\right).\label{eq:Spatial_5}
\end{align}
Let us study \eqref{eq:Spatial_5} in terms of its potential asymptotic
optimality properties. Using the triangle inequality, it is true that
\begin{multline}
\left|{\cal E}^{L_{S}}\left(\left.y_{t}\left({\bf q}\right)\right|\mathscr{Y}_{t}\right)-\widehat{y}_{t}\left({\bf q}\right)\right|\le\\
\le\left|A^{{\bf q}}\left({\cal E}^{L_{S}}\left(\left.X_{t}\right|\mathscr{Y}_{t}\right)-\mathbb{E}\left\{ \left.X_{t}\right|\mathscr{Y}_{t}\right\} \right)\right|+\\
+\left|\left({\cal E}^{L_{S}}\left(\left.\boldsymbol{\phi}_{t}^{1}\left(X_{t}\right)\right|\mathscr{Y}_{t}\right)-\mathbb{E}\left\{ \left.\boldsymbol{\phi}_{t}^{1}\left(X_{t}\right)\right|\mathscr{Y}_{t}\right\} \right)^{\boldsymbol{T}}{\bf y}_{t}\right|+\\
+\left|{\cal E}^{L_{S}}\left(\left.\phi_{t}^{2}\left(X_{t}\right)\right|\mathscr{Y}_{t}\right)-\mathbb{E}\left\{ \left.\phi_{t}^{2}\left(X_{t}\right)\right|\mathscr{Y}_{t}\right\} \right|.
\end{multline}
Also, from the Cauchy-Schwarz Inequality and the fact that the $L_{2}$
norm of a vector is upper bounded by its $L_{1}$ norm,
\begin{multline}
\left|{\cal E}^{L_{S}}\left(\left.y_{t}\left({\bf q}\right)\right|\mathscr{Y}_{t}\right)-\widehat{y}_{t}\left({\bf q}\right)\right|\le\left|\alpha^{{\bf q}}\right|\left\Vert \left({\cal E}^{L_{S}}\left(\left.X_{t}\right|\mathscr{Y}_{t}\right)-\mathbb{E}\left\{ \left.X_{t}\right|\mathscr{Y}_{t}\right\} \right)\right\Vert _{1}+\\
+\left\Vert {\bf y}_{t}\right\Vert _{2}\left\Vert {\cal E}^{L_{S}}\left(\left.\boldsymbol{\phi}_{t}^{1}\left(X_{t}\right)\right|\mathscr{Y}_{t}\right)-\mathbb{E}\left\{ \left.\boldsymbol{\phi}_{t}^{1}\left(X_{t}\right)\right|\mathscr{Y}_{t}\right\} \right\Vert _{1}+\left|{\cal E}^{L_{S}}\left(\left.\phi_{t}^{2}\left(X_{t}\right)\right|\mathscr{Y}_{t}\right)-\mathbb{E}\left\{ \left.\phi_{t}^{2}\left(X_{t}\right)\right|\mathscr{Y}_{t}\right\} \right|.
\end{multline}
Now, from (\cite{KalPetNonlinearEXTENDED2014}, Lemma 7), it follows
that for any natural $T<\infty$, there exists a bounded constant
$\gamma>1$, such that
\begin{equation}
\sup_{t\in\mathbb{N}_{T}}\left\Vert {\bf y}_{t}\left(\omega\right)\right\Vert _{2}<\sqrt{\gamma CN\left(1+\log\left(T+1\right)\right)},
\end{equation}
for all $\omega\in\widehat{\Omega}_{T}\subseteq\Omega$, with measure
at least
\begin{equation}
1-\dfrac{\exp\left(-CN\right)}{\left(T+1\right)^{CN-1}},
\end{equation}
exactly as in Theorem \ref{OUR_Filter-1}. Therefore, directly invoking
Theorems \ref{OUR_Filter-1} and \ref{Functionals}, it readily follows
that, under the respective conditions,
\begin{equation}
\lim_{L_{S}\rightarrow\infty}\sup_{t\in\mathbb{N}_{T}}\sup_{\omega\in\widehat{\Omega}_{T}}\left|{\cal E}^{L_{S}}\left(\left.y_{t}\left({\bf q}\right)\right|\mathscr{Y}_{t}\right)-\widehat{y}_{t}\left({\bf q}\right)\right|\equiv0,
\end{equation}
showing the second part of the theorem, when $\rho$, the prediction
horizon, coincides with zero. For the first part, observe that the
approximate predictor ${\cal E}^{L_{S}}\left(\left.y_{t}\left({\bf q}\right)\right|\mathscr{Y}_{t}\right)$
can be explicitly expressed as (see Theorems \ref{OUR_Filter} and
\ref{Functionals})
\begin{flalign}
{\cal E}^{L_{S}}\left(\left.y_{t}\left({\bf q}\right)\right|\mathscr{Y}_{t}\right) & =A^{{\bf q}}{\bf X}\dfrac{E_{t}}{\left\Vert E_{t}\right\Vert _{1}}+\left(\boldsymbol{\Phi}_{t}^{1}\dfrac{E_{t}}{\left\Vert E_{t}\right\Vert _{1}}\right)^{\boldsymbol{T}}{\bf y}_{t}-\left\langle \boldsymbol{\phi}_{t}^{2},\dfrac{E_{t}}{\left\Vert E_{t}\right\Vert _{1}}\right\rangle \nonumber \\
 & \equiv\left(A^{{\bf q}}{\bf X}+{\bf y}_{t}^{\boldsymbol{T}}\boldsymbol{\Phi}_{t}^{1}+\left(\boldsymbol{\phi}_{t}^{2}\right)^{\boldsymbol{T}}\right)\dfrac{E_{t}}{\left\Vert E_{t}\right\Vert _{1}}\nonumber \\
 & \equiv\left\langle \left(A^{{\bf q}}{\bf X}\right)^{\boldsymbol{T}}+\left(\boldsymbol{\Phi}_{t}^{1}\right)^{\boldsymbol{T}}{\bf y}_{t}+\boldsymbol{\phi}_{t}^{2},\dfrac{E_{t}}{\left\Vert E_{t}\right\Vert _{1}}\right\rangle ,
\end{flalign}
which, after simple algebra, can be easily shown to coincide with
the vector process $\boldsymbol{\phi}_{t}\left({\bf y}_{t}\right)$,
present in the statement of Theorem \ref{SSCP}.

In the prediction case, that is, when $\rho\ge1$, the procedure is
slightly different. Let us first define the complete filtration generated
by $X_{t+\rho},{\bf y}_{t}$ and $y_{t}\left({\bf q}\right)$ as
\begin{equation}
\left\{ \mathscr{Q}_{t}^{+\rho}\right\} _{t\in\mathbb{N}}\triangleq\left\{ \sigma\left\{ \left\{ X_{i+\rho},{\bf y}_{i},y_{i}\left({\bf q}\right)\right\} _{i\in\mathbb{N}_{t}}\right\} \right\} _{t\in\mathbb{N}}.
\end{equation}
Also, note that, for all $\rho\ge1$, the augmented observation vector
process (and therefore each one of its elements)
\begin{equation}
{\bf y}_{t+\rho}^{aug}\triangleq\begin{bmatrix}{\bf y}_{t+\rho}\\
y_{t+\rho}\left({\bf q}\right)
\end{bmatrix}\in\mathbb{R}^{\left(N+1\right)\times1}
\end{equation}
is \textit{conditionally independent of ${\bf y}_{t}^{aug},{\bf y}_{t-1}^{aug},\ldots$,
given the state at time $t+\rho$, $X_{t+\rho}$}. Thus, using the
tower property, it is true that
\begin{flalign}
\widehat{y}_{t+\rho}\left({\bf q}\right) & \equiv\mathbb{E}\left\{ \left.y_{t+\rho}\left({\bf q}\right)\right|\mathscr{Y}_{t}\right\} \nonumber \\
 & \equiv\mathbb{E}\left\{ \left.\mathbb{E}\left\{ \left.y_{t+\rho}\left({\bf q}\right)\right|\mathscr{Q}_{t}^{+\rho}\right\} \right|\mathscr{Y}_{t}\right\} \nonumber \\
 & =\mathbb{E}\left\{ \left.\mathbb{E}\left\{ \left.y_{t+\rho}\left({\bf q}\right)\right|X_{t+\rho}\right\} \right|\mathscr{Y}_{t}\right\} \nonumber \\
 & \equiv\mathbb{E}\left\{ \left.\mathbb{E}\left\{ \left.A^{{\bf q}}X_{t+\rho}+\sigma^{{\bf q}}\left(X_{t+\rho}\right)+\xi^{{\bf q}}\right|X_{t+\rho}\right\} \right|\mathscr{Y}_{t}\right\} \nonumber \\
 & =\mathbb{E}\left\{ \left.A^{{\bf q}}X_{t+\rho}\right|\mathscr{Y}_{t}\right\} ,
\end{flalign}
or, equivalently,
\begin{equation}
\widehat{y}_{t+\rho}\left({\bf q}\right)\equiv A^{{\bf q}}\mathbb{E}\left\{ \left.X_{t+\rho}\right|\mathscr{Y}_{t}\right\} .
\end{equation}
Consequently, defining the approximate spatiotemporal predictor
\begin{equation}
{\cal E}^{L_{S}}\left(\left.y_{t+\rho}\left({\bf q}\right)\right|\mathscr{Y}_{t}\right)\triangleq A^{{\bf q}}{\cal E}^{L_{S}}\left(\left.X_{t+\rho}\right|\mathscr{Y}_{t}\right),
\end{equation}
substituting ${\cal E}^{L_{S}}\left(\left.X_{t+\rho}\right|\mathscr{Y}_{t}\right)$
from Theorem \ref{OUR_Filter} and following a very similar convergence
analysis to the filtering case treated above, the respective results
present in the statement of Theorem \ref{SSCP} follow. The proof
is complete.\hfill{}\ensuremath{\blacksquare}

\bibliographystyle{IEEEtran}
\bibliography{IEEEabrv}

\begin{thebibliography}{10}
\providecommand{\url}[1]{#1}
\csname url@samestyle\endcsname
\providecommand{\newblock}{\relax}
\providecommand{\bibinfo}[2]{#2}
\providecommand{\BIBentrySTDinterwordspacing}{\spaceskip=0pt\relax}
\providecommand{\BIBentryALTinterwordstretchfactor}{4}
\providecommand{\BIBentryALTinterwordspacing}{\spaceskip=\fontdimen2\font plus
\BIBentryALTinterwordstretchfactor\fontdimen3\font minus
  \fontdimen4\font\relax}
\providecommand{\BIBforeignlanguage}[2]{{%
\expandafter\ifx\csname l@#1\endcsname\relax
\typeout{** WARNING: IEEEtran.bst: No hyphenation pattern has been}%
\typeout{** loaded for the language `#1'. Using the pattern for}%
\typeout{** the default language instead.}%
\else
\language=\csname l@#1\endcsname
\fi
#2}}
\providecommand{\BIBdecl}{\relax}
\BIBdecl

\bibitem{LiPetropuluPoor2011}
J.~Li, A.~Petropulu, and H.~Poor, ``Cooperative transmission for relay networks
  based on second-order statistics of channel state information,'' \emph{IEEE
  Transactions on Signal Processing}, vol.~59, pp. 1280 -- 1291, March 2011.

\bibitem{NikosBeam-1}
N.~Chatzipanagiotis, Y.~Liu, A.~Petropulu, and M.~Zavlanos, ``Distributed
  cooperative beamforming in multi-source multi-destination clustered
  systems,'' \emph{Signal Processing, IEEE Transactions on}, vol.~62, no.~23,
  pp. 6105--6117, Dec 2014.

\bibitem{Petropulu_1_2010}
L.~Dong, Z.~Han, A.~Petropulu, and H.~Poor, ``Improving wireless physical layer
  security via cooperating relays,'' \emph{Signal Processing, IEEE Transactions
  on}, vol.~58, no.~3, pp. 1875--1888, March 2010.

\bibitem{Petropulu_2_2008}
------, ``Secure wireless communications via cooperation,'' in
  \emph{Communication, Control, and Computing, 2008 46th Annual Allerton
  Conference on}, Sept 2008, pp. 1132--1138.

\bibitem{Petropulu_3_2009}
------, ``Cooperative jamming for wireless physical layer security,'' in
  \emph{Statistical Signal Processing, 2009. SSP '09. IEEE/SP 15th Workshop
  on}, Aug 2009, pp. 417--420.

\bibitem{Petropulu_4_2009}
------, ``Amplify-and-forward based cooperation for secure wireless
  communications,'' in \emph{Acoustics, Speech and Signal Processing, 2009.
  ICASSP 2009. IEEE International Conference on}, April 2009, pp. 2613--2616.

\bibitem{Trappe2_2009}
A.~Kaya, L.~Greenstein, and W.~Trappe, ``Characterizing indoor wireless
  channels via ray tracing combined with stochastic modeling,'' \emph{Wireless
  Communications, IEEE Transactions on}, vol.~8, no.~8, pp. 4165--4175, August
  2009.

\bibitem{CLPZHawaii2012}
N.~Chatzipanagiotis, Y.~Liu, A.~Petropulu, and M.~M. Zavlanos, ``Controlling
  groups of mobile beamformers,'' in \emph{IEEE Conference on Decision and
  Control}, Hawaii, 2012.

\bibitem{Trappe3_2008}
K.~Ma, Y.~Zhang, and W.~Trappe, ``Managing the mobility of a mobile sensor
  network using network dynamics,'' \emph{Parallel and Distributed Systems,
  IEEE Transactions on}, vol.~19, no.~1, pp. 106--120, Jan 2008.

\bibitem{KalPet-Jammers-2013}
D.~S. Kalogerias, N.~Chatzipanagiotis, M.~M. Zavlanos, and A.~P. Petropulu,
  ``Mobile jammers for secrecy rate maximization in cooperative networks,'' in
  \emph{Acoustics, Speech and Signal Processing (ICASSP), 2013 IEEE
  International Conference on}, May 2013, pp. 2901--2905.

\bibitem{KalPet-Mobi-2014}
D.~S. Kalogerias and A.~P. Petropulu, ``Mobi-cliques for improving ergodic
  secrecy in fading wiretap channels under power constraints,'' in
  \emph{Acoustics, Speech and Signal Processing (ICASSP), 2014 IEEE
  International Conference on}, May 2014, pp. 1578--1591.

\bibitem{Mostofi_3_2011}
A.~Ghaffarkhah and Y.~Mostofi, ``Communication-aware motion planning in mobile
  networks,'' \emph{Automatic Control, IEEE Transactions on}, vol.~56, no.~10,
  pp. 2478--2485, Oct 2011.

\bibitem{Mostofi_4_2012}
Y.~Yan and Y.~Mostofi, ``Robotic router formation in realistic communication
  environments,'' \emph{Robotics, IEEE Transactions on}, vol.~28, no.~4, pp.
  810--827, Aug 2012.

\bibitem{Mostofi_5_2014}
A.~Ghaffarkhah and Y.~Mostofi, ``Dynamic networked coverage of time-varying
  environments in the presence of fading communication channels,'' \emph{ACM
  Transactions on Sensor Networks (TOSN)}, vol.~10, no.~3, p.~45, 2014.

\bibitem{Multipath_1_2011}
D.~Shutin and B.~Fleury, ``Sparse variational bayesian sage algorithm with
  application to the estimation of multipath wireless channels,'' \emph{Signal
  Processing, IEEE Transactions on}, vol.~59, no.~8, pp. 3609--3623, Aug 2011.

\bibitem{EM_SAGE_1994}
J.~Fessler and A.~Hero, ``Space-alternating generalized
  expectation-maximization algorithm,'' \emph{Signal Processing, IEEE
  Transactions on}, vol.~42, no.~10, pp. 2664--2677, Oct 1994.

\bibitem{Multipath_2_2009}
J.~Salmi, A.~Richter, and V.~Koivunen, ``Detection and tracking of mimo
  propagation path parameters using state-space approach,'' \emph{Signal
  Processing, IEEE Transactions on}, vol.~57, no.~4, pp. 1538--1550, 2009.

\bibitem{Multipath_3_2008}
X.~Yin, G.~Steinbock, G.~Kirkelund, T.~Pedersen, P.~Blattnig, A.~Jaquier, and
  B.~Fleury, ``Tracking of time-variant radio propagation paths using particle
  filtering,'' in \emph{Communications, 2008. ICC '08. IEEE International
  Conference on}, May 2008, pp. 920--924.

\bibitem{Giannakis_Spatial1_2011}
S.-J. Kim, E.~Dall'Anese, and G.~Giannakis, ``Cooperative spectrum sensing for
  cognitive radios using kriged kalman filtering,'' \emph{Selected Topics in
  Signal Processing, IEEE Journal of}, vol.~5, no.~1, pp. 24--36, Feb 2011.

\bibitem{Giannakis_Spatial2_2011}
E.~Dall'Anese, S.-J. Kim, and G.~Giannakis, ``Channel gain map tracking via
  distributed kriging,'' \emph{Vehicular Technology, IEEE Transactions on},
  vol.~60, no.~3, pp. 1205--1211, March 2011.

\bibitem{Channel_Modeling1_2009}
P.~Agrawal and N.~Patwari, ``Correlated link shadow fading in multi-hop
  wireless networks,'' \emph{Wireless Communications, IEEE Transactions on},
  vol.~8, no.~8, pp. 4024--4036, August 2009.

\bibitem{KKF_1_1999}
C.~K. Wikle and N.~Cressie, ``A dimension-reduced approach to space-time kalman
  filtering,'' \emph{Biometrika}, vol.~86, no.~4, pp. 815--829, 1999.

\bibitem{KKF_2_1998}
K.~V. Mardia, C.~Goodall, E.~J. Redfern, and F.~J. Alonso, ``The kriged kalman
  filter,'' \emph{Test}, vol.~7, no.~2, pp. 217--282, 1998.

\bibitem{KKF_3_2009}
J.~Cortes, ``Distributed kriged kalman filter for spatial estimation,''
  \emph{Automatic Control, IEEE Transactions on}, vol.~54, no.~12, pp.
  2816--2827, Dec 2009.

\bibitem{MostofiSpatial2012}
M.~Malmirchegini and Y.~Mostofi, ``On the spatial predictability of
  communication channels,'' \emph{Wireless Communications, IEEE Transactions
  on}, vol.~11, no.~3, pp. 964--978, March 2012.

\bibitem{Elliott1994Hidden}
R.~J. Elliott, L.~Aggoun, and J.~B. Moore, \emph{Hidden Markov Models}.\hskip
  1em plus 0.5em minus 0.4em\relax Springer, 1994.

\bibitem{Goldsmith2005Wireless}
A.~Goldsmith, \emph{Wireless communications}.\hskip 1em plus 0.5em minus
  0.4em\relax Cambridge university press, 2005.

\bibitem{Cotton2007}
S.~Cotton and W.~Scanlon, ``Higher order statistics for lognormal small-scale
  fading in mobile radio channels,'' \emph{Antennas and Wireless Propagation
  Letters, IEEE}, vol.~6, pp. 540--543, 2007.

\bibitem{Gudmundson1991}
M.~Gudmundson, ``Correlation model for shadow fading in mobile radio systems,''
  \emph{Electronics Letters}, vol.~27, no.~23, pp. 2145--2146, Nov 1991.

\bibitem{Segall_Point1976}
A.~Segall, ``Recursive estimation from discrete-time point processes,''
  \emph{Information Theory, IEEE Transactions on}, vol.~22, no.~4, pp.
  422--431, Jul 1976.

\bibitem{Marcus1979}
S.~Marcus, ``Optimal nonlinear estimation for a class of discrete-time
  stochastic systems,'' \emph{Automatic Control, IEEE Transactions on},
  vol.~24, no.~2, pp. 297--302, Apr 1979.

\bibitem{Elliott1994Exact}
R.~J. Elliott, ``Exact adaptive filters for markov chains observed in gaussian
  noise,'' \emph{Automatica}, vol.~30, no.~9, pp. 1399--1408, 1994.

\bibitem{Elliott1994_HowToCount}
R.~J. Elliott and H.~Yang, ``How to count and guess well: Discrete adaptive
  filters,'' \emph{Applied Mathematics and Optimization}, vol.~30, no.~1, pp.
  51--78, 1994.

\bibitem{Segall1976}
A.~Segall, ``Stochastic processes in estimation theory,'' \emph{Information
  Theory, IEEE Transactions on}, vol.~22, no.~3, pp. 275--286, May 1976.

\bibitem{KalPetGRID2014}
D.~S. Kalogerias and A.~P. Petropulu, ``Grid-based filtering of {M}arkov
  processes revisited: Asymptotic optimality \& recursive estimation,''
  \emph{IEEE Transactions on Signal Processing}, Submitted in 2015. Available
  at:
  {\href{http://eceweb1.rutgers.edu/~dkalogerias/Grid_Filtering_2015.pdf}{Link}}.

\bibitem{Pages2005optimal}
G.~Pag{\`e}s, H.~Pham \emph{et~al.}, ``Optimal quantization methods for
  nonlinear filtering with discrete-time observations,'' \emph{Bernoulli},
  vol.~11, no.~5, pp. 893--932, 2005.

\bibitem{Papoulis2002}
A.~Papoulis and S.~U. Pillai, \emph{Probability, Random Variables and
  Stochastic Processes}, 4th~ed.\hskip 1em plus 0.5em minus 0.4em\relax
  McGraw-Hill, 2002.

\bibitem{PARTICLE2002tutorial}
M.~S. Arulampalam, S.~Maskell, N.~Gordon, and T.~Clapp, ``A tutorial on
  particle filters for online nonlinear/non-gaussian bayesian tracking,''
  \emph{Signal Processing, IEEE Transactions on}, vol.~50, no.~2, pp. 174--188,
  2002.

\bibitem{Gordon1993Novel}
N.~J. Gordon, D.~J. Salmond, and A.~F. Smith, ``Novel approach to
  nonlinear/non-gaussian bayesian state estimation,'' in \emph{IEE Proceedings
  F (Radar and Signal Processing)}, vol. 140, no.~2.\hskip 1em plus 0.5em minus
  0.4em\relax IET, 1993, pp. 107--113.

\bibitem{Crisan2002Survey}
D.~Crisan and A.~Doucet, ``A survey of convergence results on particle
  filtering methods for practitioners,'' \emph{Signal Processing, IEEE
  Transactions on}, vol.~50, no.~3, pp. 736--746, 2002.

\bibitem{VikramDoucet2001Particle}
A.~Doucet, N.~J. Gordon, and V.~Krishnamurthy, ``Particle filters for state
  estimation of jump markov linear systems,'' \emph{Signal Processing, IEEE
  Transactions on}, vol.~49, no.~3, pp. 613--624, 2001.

\bibitem{DAUM_CURSE2003}
F.~Daum and J.~Huang, ``Curse of dimensionality and particle filters,'' in
  \emph{Aerospace Conference, 2003, Proceedings IEEE}, vol.~4, March 2003, pp.
  1979--1993.

\bibitem{Bengtsson2008Curse}
T.~Bengtsson, P.~Bickel, B.~Li \emph{et~al.}, ``Curse-of-dimensionality
  revisited: Collapse of the particle filter in very large scale systems,''
  \emph{Probability and Statistics: Essays in Honor of David A. Freedman},
  vol.~2, pp. 316--334, 2008.

\bibitem{Quang2010Insight}
P.~B. Quang, C.~Musso, and F.~Le~Gland, ``An insight into the issue of
  dimensionality in particle filtering,'' in \emph{Information Fusion (FUSION),
  2010 13th Conference on}.\hskip 1em plus 0.5em minus 0.4em\relax IEEE, 2010,
  pp. 1--8.

\bibitem{RebeschiniRamon2013_1}
P.~Rebeschini and R.~van Handel, ``Can local particle filters beat the curse of
  dimensionality?'' \emph{http://arxiv.org/abs/1301.6585}, 2013.

\bibitem{KalPetNonlinearEXTENDED2014}
D.~S. Kalogerias and A.~P. Petropulu, ``Asymptotically optimal discrete time
  nonlinear filters from stochastically convergent state process
  approximations,'' \emph{IEEE Transactions on Signal Processing}, Submitted in
  2015. Extended version available at:
  {\href{http://arxiv.org/abs/1411.6719}{Link}}.

\end{thebibliography}

\end{document}